\documentclass{ar-1col}
\usepackage[comma]{natbib}
\usepackage{amsmath,amssymb}
\usepackage{mathtools}
\usepackage{url}
\usepackage{hyperref}
\usepackage{pdflscape}
\usepackage[version=4]{mhchem}

\jname{ Annual Review of Astronomy and Astrophysics}
\jvol{61}
\jyear{2022}
\begin{document}
\markboth{Jewitt \& Seligman}{Interstellar Interlopers}
\title{The Interstellar Interlopers }
\author{David Jewitt$^1$, and Darryl Z. Seligman$^2$ 
\affil{$^1$Department of Earth, Planetary and Space Sciences, UCLA, 595 Charles Young Drive East, Los Angeles, CA 90095-1567; email: jewitt@ucla.edu; ORCID: 0000-0003-0262-8160}
\affil{$^2$Department of Astronomy and Carl Sagan Institute, Cornell University, 122 Sciences Drive, Ithaca, NY, 14853, USA; email: dzs9@cornell.edu; ORCID: 0000-0002-0726-6480}}

\begin{abstract}
Interstellar interlopers are bodies formed outside of the solar system but observed passing through it. The first  two identified interlopers, 1I/`Oumuamua and 2I/Borisov, exhibited unexpectedly different physical properties. 1I/`Oumuamua appeared unresolved and asteroid-like whereas  2I/Borisov was a more comet-like source of both gas and dust. Both objects moved under the action of non-gravitational acceleration. These interlopers and their divergent properties provide our only window so far onto an enormous and previously unknown galactic population.  The number density of such objects is $\sim$ 0.1 AU$^{-3}$ which, if uniform across the galactic disk, would imply 10$^{25}$ to 10$^{26}$ similar objects in the Milky Way.  The interlopers likely formed in, and were ejected from, the protoplanetary disks of young stars. However, we currently possess too little data to firmly reject other explanations.

\begin{itemize}
\item 1I/`Oumuamua and 2I/Borisov are gravitationally unbound, \\sub-kilometer bodies both showing non-gravitational acceleration. 
\item The acceleration of 1I/`Oumuamua in the absence of measurable \\mass loss   requires  either a strained explanation in terms of recoil from \\sublimating super-volatiles or the action of radiation pressure on a \\nucleus  with an ultra-low mass column density, $\sim$ 1 kg m$^{-2}$.  
\item 2I/Borisov is a strong source of CO and H$_2$O, which together account \\for its activity and non-gravitational acceleration. 
\item The interlopers are most likely planetesimals from the protoplanetary \\disks of other  stars, ejected by gravitational scattering from planets. \\ 1I/`Oumuamua and 2I/Borisov have dynamical ages $\sim10^8$ and \\$\sim10^9$ years, respectively.  
\item Forthcoming observatories should detect interstellar interlopers \newline every year, which will provide a rapid boost to our knowledge \\ 
of the population.
\end{itemize}
\end{abstract}
\maketitle
\tableofcontents
\section{Background}
Tiny solid particles of dust carry about 1\% of the mass of the interstellar medium.  Formed in the expanding, unstable atmospheres of  evolved stars, these particles  have  characteristic sizes $\sim$0.2 to 0.3 $\mu$m or less. Until recently, the existence, abundance, and nature of possible macroscopic (meter-sized and larger)  interstellar bodies has been a matter  for speculation only.  With the discovery of  1I/`Oumuamua and 2I/Borisov, two sub-kilometer bodies passing through the solar system but formed elsewhere, the new population of macroscopic interstellar interlopers has finally been revealed. 

The large sizes of the interlopers imply formation in dense environments, presumably the protoplanetary disks where interstellar dust hierarchically accumulates  to  form planets. Gravitational scattering by growing planets can excite the velocity dispersion of planetesimals, launching some beyond the control of the parent star. Judging from our own solar system, most of the mass of the protoplanetary disk, and most of the material ejected to the interstellar medium by planetary scattering, originated in the cold regions beyond the H$_2$O snowline. This freezeout line is currently close to the orbit of Jupiter in our solar system.  The ice-rich material draws an immediate parallel between the expected properties of the interstellar bodies and those of the comets native to the Solar System, which we briefly discuss here. 

Solar system comets are divided into two distinct dynamical types corresponding to two distinct long-term storage reservoirs. The short-period comets (SPCs), with modest eccentricities ($e <$ 1) and inclinations, are derived from the trans-Neptunian Kuiper belt  (specifically, from the scattered disk component of the Kuiper belt) by long-term chaotic instabilities \citep{Volk08}.  The mass of the present-day Kuiper belt is $\sim$0.1 M$_{\oplus}$ (reduced from an estimated starting mass of $\sim$20 M$_{\oplus}$ or 30 M$_{\oplus}$). The belt contains  $\sim10^5$ bodies larger than 100 km and perhaps $\sim 10^{10}$  of kilometer-size.  Some Kuiper belt objects (the so-called ``Cold Classical'' objects) likely formed in-situ \citep{Parker2010,Nesvorny2018}, while other components of the belt are suspected to have formed at smaller distances and were subsequently scattered out gravitationally.

\begin{figure}
\begin{center}
       \includegraphics[scale=0.32,angle=0]{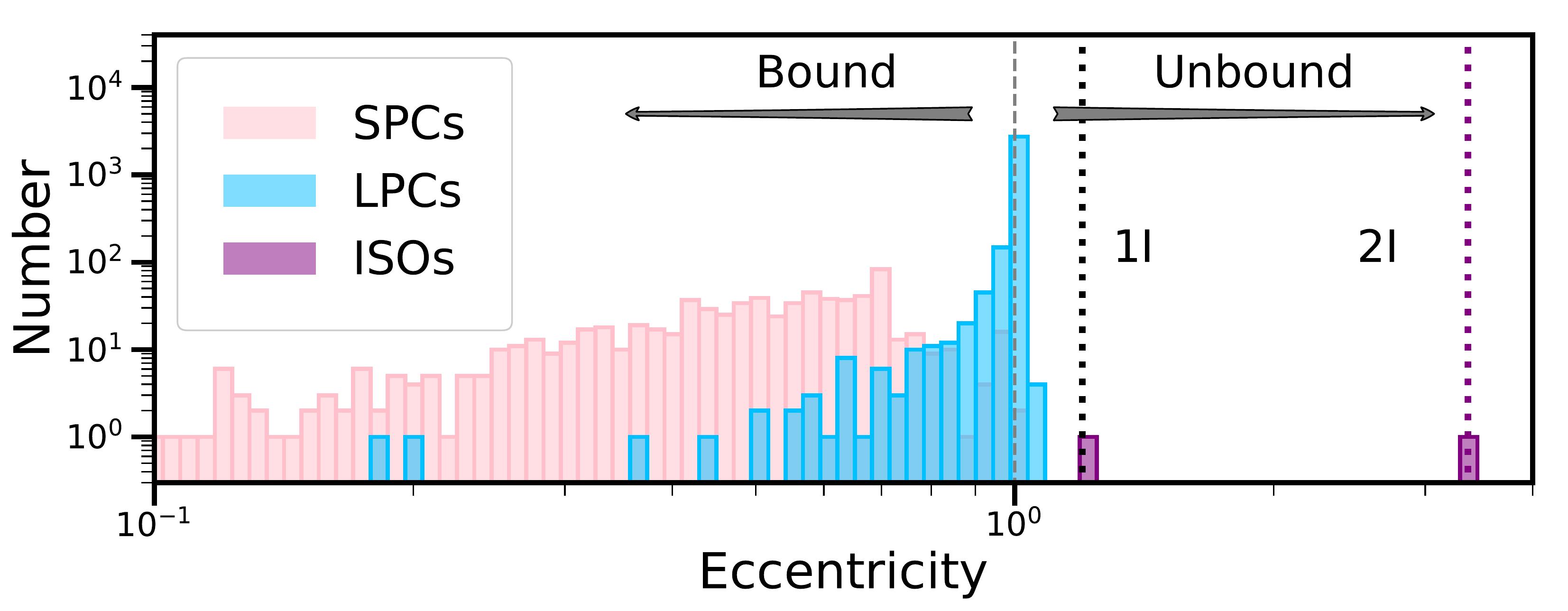}
    \caption{The distribution of osculating orbital eccentricities of LPCs (blue), SPCs (pink) and ISOs (purple).  The dashed grey line indicates  $e=1$, separating  bound from unbound orbits. The LPCs cluster at $e\simeq1$. The eccentricities of 1I and 2I are indicated with black and purple dotted lines, respectively.  }\label{fig:orbital_elements} 
\end{center}
\end{figure}

In contrast, the long-period comets (LPCs) have an isotropic distribution of inclinations,  large semimajor axes, and eccentricities, $e \sim$ 1 (Figure \ref{fig:orbital_elements}).  Their source is the Oort cloud, a 50,000 AU scale gravitationally bound swarm estimated to contain $N_{OC} \sim10^{11}$ \citep{francis2005} to 10$^{12}$ \citep{Brasser13, Dones15} kilometer-sized comets, as first recognized by  \citet{Oort1950}. Although values for the total mass of the Oort cloud $\sim$1 M$_{\oplus}$ are  widely quoted, masses up to $\sim$20 M$_{\oplus}$ \citep{francis2005} are allowed by the data. This is because the sizes and size distribution of LPC nuclei are poorly known, and because the inner Oort cloud (semimajor axes $\lesssim$few $\times10^3$ AU)  might contain substantial mass and yet go undetected, because it is relatively immune to external perturbations (but see \cite{Kaib09}).  Even during the earliest stages of planetary formation, typical densities at Oort cloud  distances were too low for comets to have formed in situ there.  Instead, the LPCs likely formed in the region now occupied by the giant planets, whose growth lead to the gravitational scattering of many comets into highly eccentric and unbound ($e >$ 1) orbits. A fraction of those ejected comets, 0.01 $\lesssim f \lesssim$ 0.1  \citep{Hahn99, Brasser10}, became captured (by external perturbations from nearby stars, perhaps in the Sun's birth cluster \citep{Dones15}, and by the galactic tide) into  circularized orbits with raised perihelia. Once removed from the gravitational control of the planets, the inclinations of these captured orbits were randomized over the course of several Gyr, transforming the initial disk-like distribution into the spherical, modern-day Oort cloud \citep{Higuchi15}. Continuing external perturbations on the Oort cloud supply  LPCs back to the planetary region, while dispersing others to  interstellar space.  

The remaining $(1-f)\simeq$ 90\% - 99\% of the ejected objects were lost by the Sun to the interstellar medium. For example, taking $f$ = 0.01, as many as  $N_{OC}/f \sim$ 10$^{13}$ to 10$^{14}$ comets were ejected from the solar system alone.    Given $N_{\star} \sim10^{11}$ stars in the galaxy, and assuming that all stars have Sun-like comet populations, a first order guess as to the number of ejected comets in the galaxy  is $N_{ii} = N_{\star} N_{OC}/f \sim$ 10$^{24}$ to 10$^{25}$. The  combined  mass of this population is a large but very uncertain \textbf{$\sim$}10$^6$ M$_{\odot}$ to \textbf{$\sim$}10$^7$ M$_{\odot}$ (assuming a nominal 1 km radius, as is typical of the measured comets). Depending on the nucleus size distribution, the total ejected mass could be much larger, as could the  numbers of ejected sub-kilometer comets.

These simplified estimates are necessarily extrapolations based on limited knowledge.  As discussed previously, the fraction of scattered comets that remained bound to the Solar System is  uncertain by an order of magnitude. Moreover, the analogous fraction for other stars depends sensitively on the architecture of their planetary systems and on the star cluster environment when they formed.  The ejection and capture efficiencies must therefore  differ from star to star, perhaps by large factors.  Nevertheless, the serendipitous discoveries of 1I/`Oumuamua and 2I/Borisov prove the existence of a vast galactic population of macroscopic interlopers beyond reasonable doubt.  There is explosive interest in both the physical nature and the possible origins of the galactic reservoir of interstellar interlopers.  
\begin{figure}
\begin{center}
       \includegraphics[scale=0.3,angle=0]{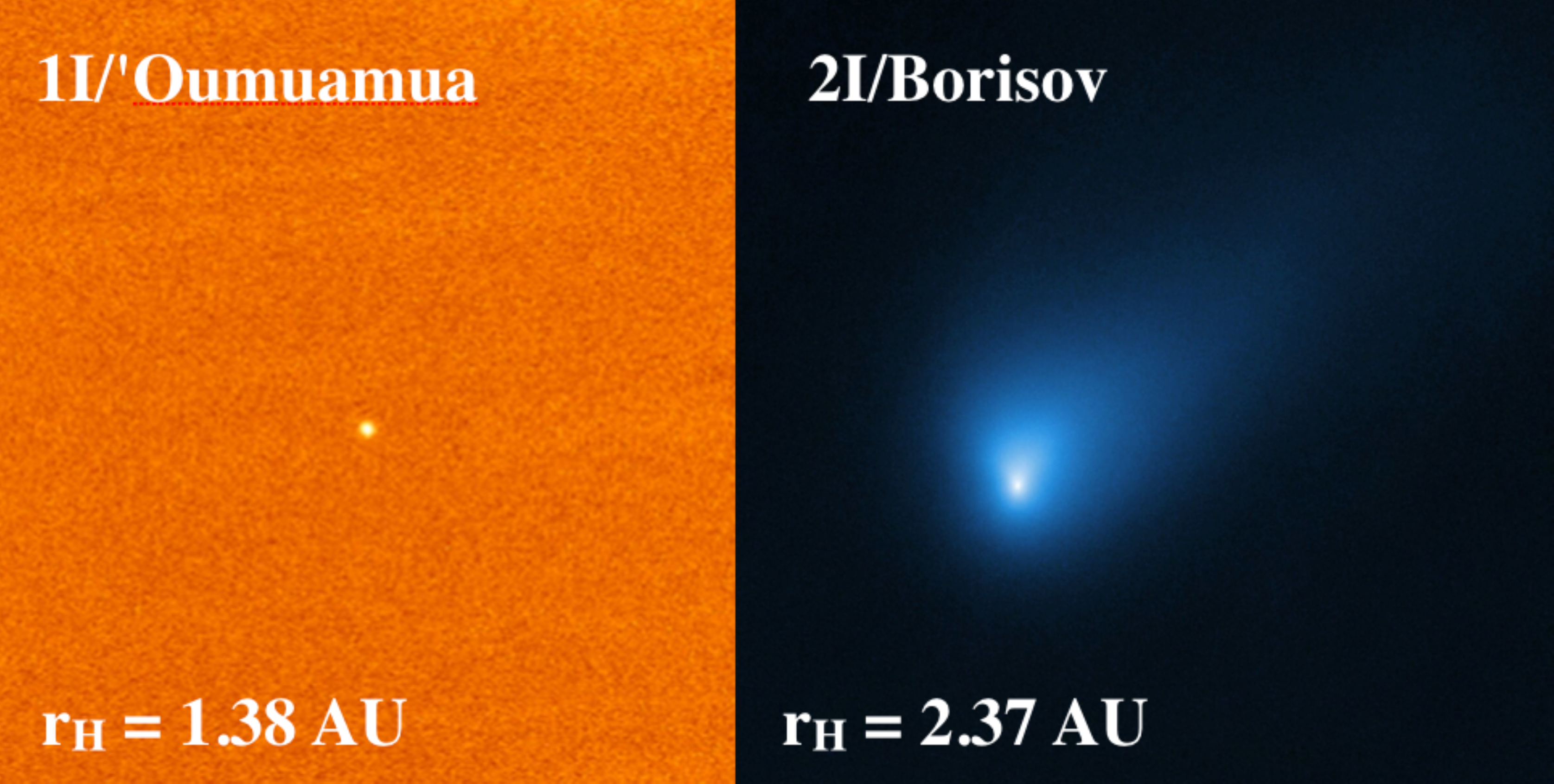}
    \caption{(left:) 1I/`Oumuamua showing point-like appearance on UT 2017 October 26 at the 2.5 m Nordic Optical Telescope (right:) 2I/Borisov showing cometary activity on UT 2019 October 12 from the 2.4 m Hubble Space Telescope. }
 \label{1I_and_2I} 
\end{center}
\end{figure}
 \section{Dynamical Properties}
The basic dynamical properties of the first two interstellar interlopers are summarized in Table \ref{dynamical_properties} while their appearances at optical wavelengths are shown in Figure \ref{1I_and_2I}.

\begin{table}
\tabcolsep1.5pt
\caption{Dynamical Properties of the Interstellar Interlopers$^{a,b}$}
\label{dynamical_properties}
\begin{center}
\begin{tabular}{|@{}c|c|c|c|c|c|c|c|c|c|c|c@{}}
\hline
Object & T$_P$ & a  & e & i  & q  & $10^8 A_1$  & $10^8 A_2$   & $10^8 A_3$   & $\alpha_{ng}(1)$            \\
 & &  [AU] &  &  [deg] &  [AU] & AU d$^{-2}$  & AU d$^{-2}$   & AU d$^{-2}$   &  [m s$^{-2}$]    \\
\hline
1I & 2017-Sep-09 & -1.272 & 1.198 & 122.8 & 0.252 &  27.9 &  1.4   &  1.6 & 5.6$\times10^{-6}$ \\\hline
2I & 2019-Dec-08 & -0.850 & 3.363 & 44.0 & 2.009 &  7.1 &  -1.4  &  0.1 & 1.5$\times10^{-6}$  \\\hline
\end{tabular}
\end{center}
\begin{tabnote}
$^{\rm a}$ T$_P$ is the date of perihelion, $a$, $e$, $i$, $q$ are barycentric pre-entry orbital elements computed for 1900 January 1. $a < 0$ denotes hyperbolic orbit.\\
$^{\rm b}$ $A_1$, $A_2$, $A_3$, and $\alpha_{ng}(1)$ are non-gravitational acceleration parameters described in \S 2.2. 
\end{tabnote}
\end{table}

 \subsection{Discovery and Orbits}\label{Sec:orbits}
 
\begin{figure}
\begin{center}
\includegraphics[scale=0.5,angle=0]{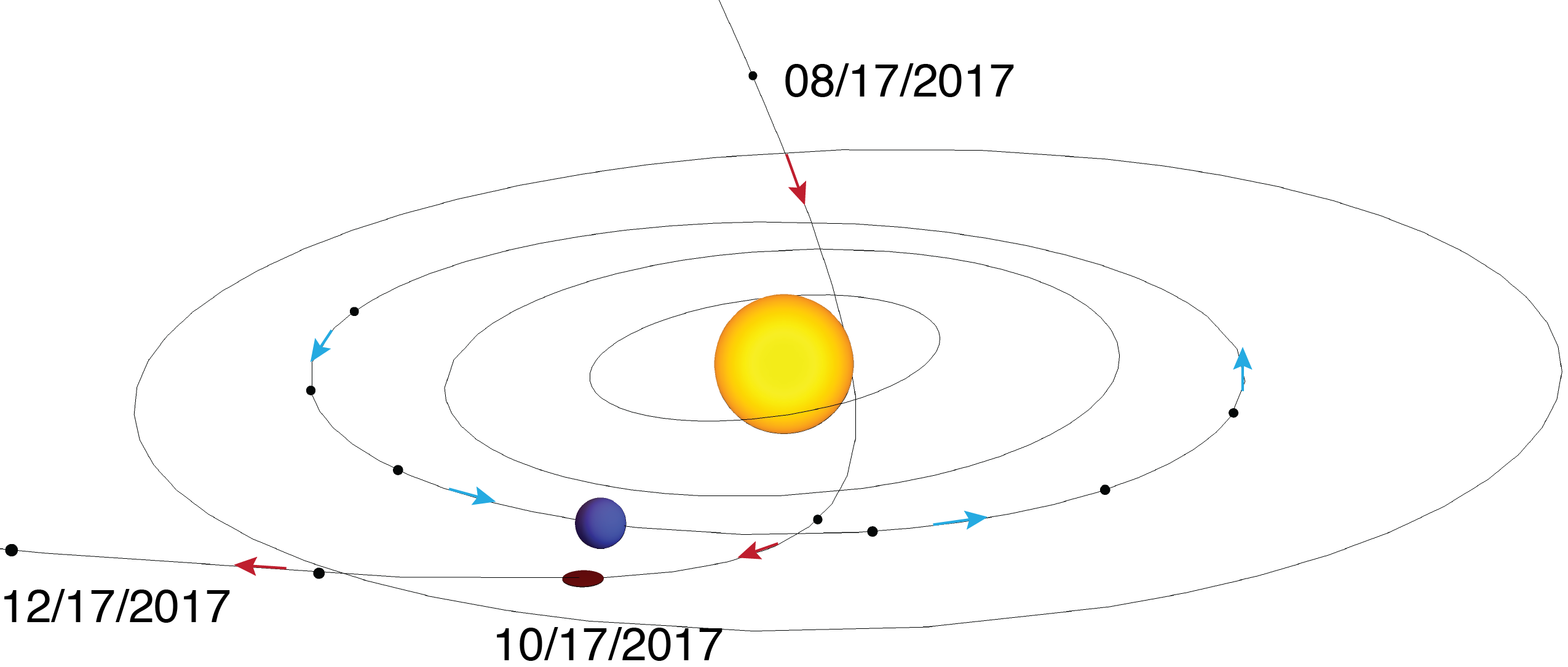}
    \caption{The trajectory of 1I/`Oumuamua near perihelion, from 2017 August to December. The positions of the Earth and `Oumuamua at discovery on October 17 are indicated. Even black points indicate the positions of the Earth and `Oumuamua at time intervals of one month. Arrows indicate the directions of motion of both objects through their orbits. 
}\label{fig:orbit_1I} 
\end{center}
\end{figure}

 \textbf{1I/`Oumuamua} The first confirmed interstellar object (formerly C/2017 U1) was discovered by Robert Weryk on 2017 October 19 from the summit of   Haleakalā on Maui, Hawaii (announced as 2017 U1 in \cite{Williams17}).  Observations soon revealed that the orbit was hyperbolic. Specifically, the object had  an osculating eccentricity $e =$ 1.201, retrograde inclination, $i$ = 122.8 deg, and a perihelion distance $q =$ 0.256 AU. (Removal of planetary perturbations gives only slightly different pre-entry barycentric orbital elements, as listed in Table \ref{dynamical_properties}). Figure \ref{fig:orbit_1I}, shows the trajectory of `Oumuamua and its position on 2017 October 17. `Oumuamua was discovered close to the Earth, passing within only about 0.16 AU three days prior to discovery.  This serendipitous detection near the peak of ground-based visibility implies the existence of an unnoticed population of similar objects with less favorable observing geometries.  1I/`Oumuamua was point-like in all observations (Figure \ref{1I_and_2I}).

\begin{figure}
\begin{center}
\includegraphics[scale=0.6,angle=0]{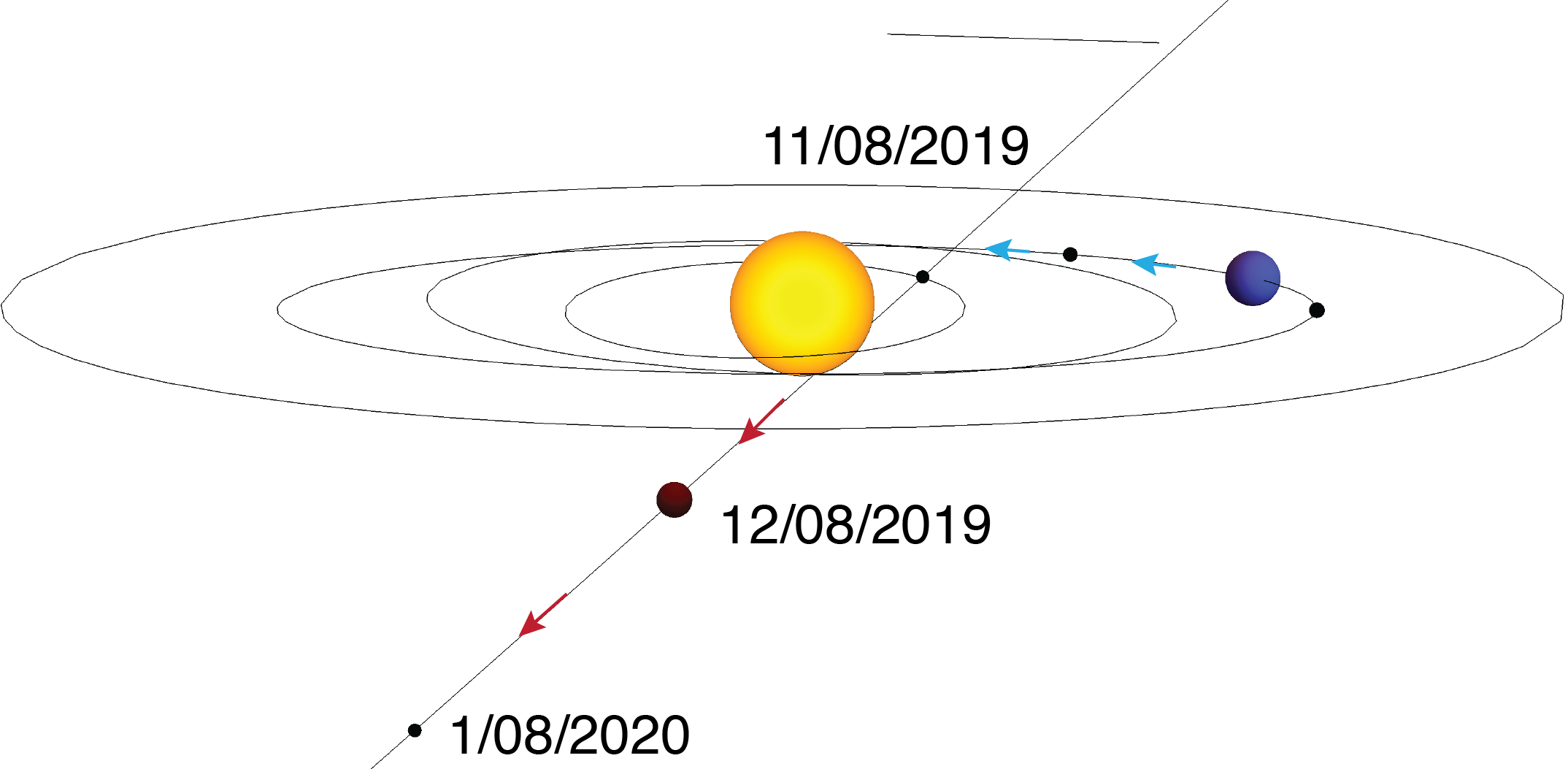}
    \caption{The trajectory of 2I/Borisov. The positions of the Earth and 2I/Borisov at discovery are marked, with month-separated points  as in Figure \ref{fig:orbit_1I}. }
\label{fig:orbit_2I} 
\end{center}
\end{figure}

 \textbf{2I/Borisov:}  The second known interstellar interloper, 2I/Borisov (also known as C/2019 Q4), was discovered on UT 2019 August 30 when only 38$^{\circ}$~from the Sun. This remarkable observation was made by Gennadiy Borisov who used a 0.65 m self-built telescope to target an area of the sky barely examined by other survey telescopes. Its  orbit is robustly hyperbolic with  eccentricity $e = 3.358$, and consequently the interstellar origin of 2I is not in doubt.  `Oumuamua was discovered outbound from perihelion only by virtue of its close approach to Earth, and it was only observable for a short time-span.  In sharp contrast -- and fortunately --   2I/Borisov was discovered about three months before perihelion (UT 2019 December 8 at $q$ = 2.00 AU) because of its bright coma  of ejected dust (Figure \ref{1I_and_2I}).  The early discovery and intrinsic brightness of 2I/Borisov enabled physical and astrometric observations to be obtained for about  1 year, whereas `Oumuamua faded beyond detection in only 2.5 months.  The trajectory of 2I/Borisov is shown schematically in Figure \ref{fig:orbit_2I}. 
 
\subsection{Non-Gravitational Acceleration}
\begin{figure}
\begin{center}
       \includegraphics[scale=0.25,angle=0]{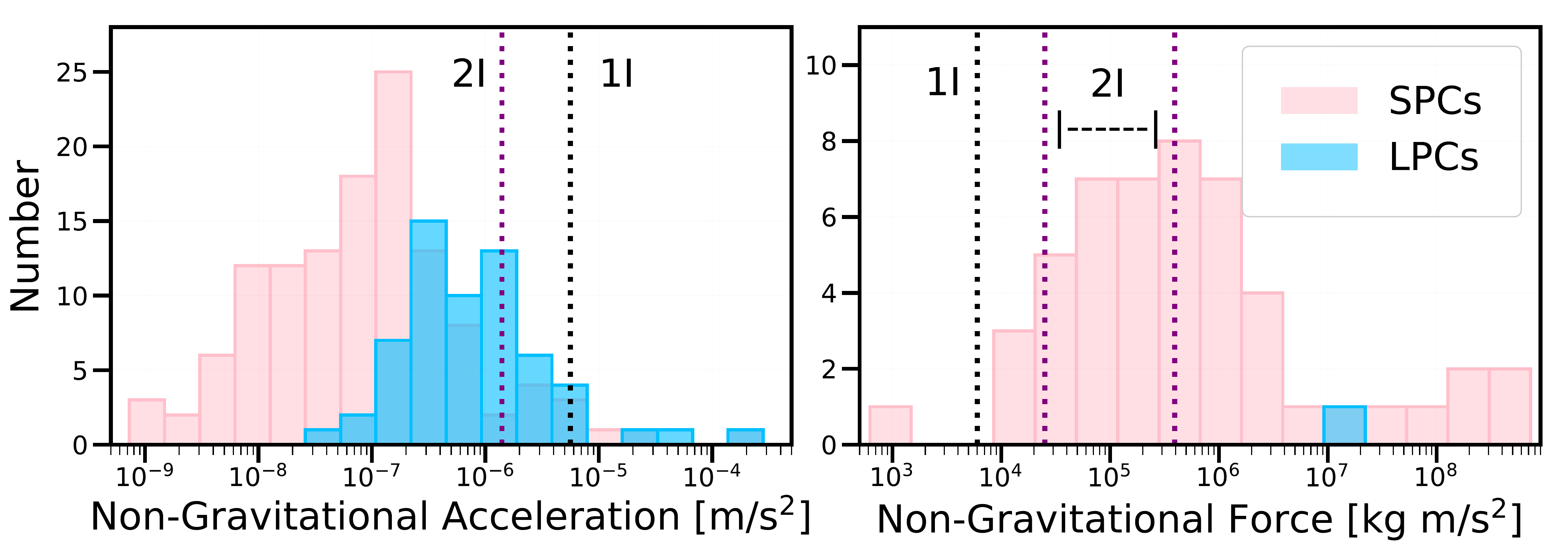}
    \caption{ (left panel) Dashed lines show the measured non-gravitational accelerations of 1I/`Oumuamua and 2I/Borisov compared to values for short-period (pink) and long-period (blue) comets.    (right panel) Non-gravitational force computed assuming spherical nuclei and a bulk density  $\rho=500$ kg/m$^3$. For the interlopers, we use  object radii as listed in Table (\ref{physical_properties}); 2I is represented by a range to reflect the uncertainty of the radius. Non-gravitational accelerations and nucleus diameters are taken from  the  JPL Small Body Database. The sizes (hence, masses) of long-period comets are largely unknown.}\label{fig:nongravs} 
\end{center}
\end{figure}

   \textbf{1I/`Oumuamua:} Observations over a 2.5 month arc allowed an accurate determination of the orbit of `Oumuamua and unexpectedly revealed the existence of non-gravitational acceleration.  In active comets, non-gravitational acceleration is caused by recoil from the asymmetric ejection of mass.  Its detection in 1I/`Oumuamua was especially puzzling because observations provided no evidence for mass loss at any level.  By convention, such accelerations are  described by   three orthogonal components, $A_1, A_2, A_3$, each expressed as AU day$^{-2}$. Here, $A_1$ is the acceleration away from the Sun, $A_3$ is perpendicular to the orbit plane and $A_2$ completes the right-handed triple. The total acceleration at $r_H$ = 1 AU is computed from $\alpha_{ng}(1) = g(r_H) (A_1^2 + A_2^2 + A_3^3)^{1/2}$ and is expressed in m s$^{-2}$ (see Table \ref{dynamical_properties}).  The quantity $g(r_H)$ is a function to represent the radial dependence of the mass loss, normalized such that $g(1$AU) = 1 \citep{Marsden1973}. Table \ref{dynamical_properties} lists the accelerations provided by JPL Horizons\footnote{\url{https://ssd.jpl.nasa.gov/horizons/app.html}} and shows that 
 $A_1$ is the dominant component. This is consistent with a force acting radially away from the Sun, as expected of outgassing recoil (see Section \ref{recoil}) and radiation pressure (Section \ref{radiation}).  The magnitude of the acceleration, $\alpha_{ng}(1)$, corresponds to about $10^{-3}$ times the local solar gravitational acceleration at 1 AU, and is three to four orders of magnitude larger than typical accelerations in solar system comets (Figure \ref{fig:nongravs}).  However, when considering the relatively small size of `Oumuamua, the non-gravitational force is lower than those observed in typical comets. 
 
   \textbf{2I/Borisov:}  A 10 month astrometric series also revealed non-gravitational acceleration  in 2I/Borisov, with $A_1$ again being the dominant component and with a smaller magnitude (Table \ref{dynamical_properties}).  Unlike the case of 1I/`Oumuamua, the non-gravitational acceleration of 2I/Borisov was matched by obvious, on-going mass loss and the cause is not mysterious. 

Figure \ref{fig:nongravs} compares the non-gravitational accelerations and forces acting on the comets and interlopers.   Comparison of the left and right panels shows that, while the acceleration of `Oumuamua is larger than in most comets, this is simply because it is very small (acceleration is $\propto r_n^{-1}$).    When plotted as force, 1I is on the weaker side of the comet distribution, consistent with a low rate of mass loss.  The larger nucleus of 2I/Borisov shows a high but less extreme value of $\alpha_{ng}(1)$, but is unremarkable on the histogram of forces in the figure.  In both interlopers, $\alpha_{ng}$ varies with heliocentric distance approximately as $r_H^{-2}$, although the heliocentric dependence is not well measured because of the limited range of distances over which astrometric observations were secured.

\section{Physical Properties}
Key physical properties of `Oumuamua and 2I/Borisov are  summarized in Table \ref{physical_properties} and more fully discussed in the following text.

\begin{table}
\tabcolsep1.5pt
\caption{Physical Properties of the Interstellar Interlopers$^a$ }
\label{physical_properties}
\begin{center}
\begin{tabular}{|@{}c|c|c|c|c|c|c|c|c|c|c|c@{}}
\hline
Object &   Radius, $r_n^{(b)}$ & Radius, $r_n^{(c)}$ & Period, P$^{(d)}$ & Shape &$b/a^e$ & $S'^{(f)}$ & $\dot{M}^{(g)}$   \\
    &    [m]  & [m] & [hour] & &  & [\%/1000\AA] & [kg s$^{-1}$]\\
\hline
1I & 55 - 114 & 80 & $\sim$8 & Oblate & $\sim$6:1 & 15$\pm$5 & $<$10$^{-3}$\\
\hline
2I & 200 - 500 & 400 & ? & ? &  ? & 12$\pm$1 & $\sim$20 to 40\\
\hline
\end{tabular}
\end{center}
\begin{tabnote}
$^{\rm a}$Approximate physical properties as described in the text, 
$^{\rm b}$range of reported radii, expressed as the radius of an equal-area circle, albedo $p$ = 0.1 assumed,
$^{\rm c}$nominal radius adopted in this work,
$^{\rm d}$nucleus rotation period, 
$^{\rm e}$ratio of the nucleus semi-axes, 
$^{\rm f}$reflectivity gradient (Sun = 0), 
$^{\rm g}$measured mass loss rate \\
\end{tabnote}
\end{table}

\subsection{Nucleus Sizes}

The apparent magnitude of an object viewed in reflected sunlight, $V$, is related  to the albedo, $p$, and effective radius, $r_n$ by \citep{Russell16}

\begin{equation}
p \,\Phi(\alpha) \, r_n^2 = \,\bigg(\,2.25\times10^{22} \,\bigg)\,r_H^2\, \Delta^2 \,10^{0.4(V_{\odot} - V)}\,,
\label{russel}
\end{equation}

\noindent where $r_H$ and $\Delta$ are the heliocentric and geocentric distances in AU, $\alpha$ is the phase (Sun-object-observer) angle, and $V_{\odot}$ is the apparent magnitude of the Sun.  The  phase function, $\Phi(\alpha)$, represents the dimming of the object observed at phase angle $\alpha$ relative to  $\alpha$ = 0.  Given only optical observations, the number of unknowns ($p$, $r_n$ and $\Phi(\alpha)$) in Equation \ref{russel} exceeds the number of constraints. Therefore, there is  ambiguity in estimates of $r_n$ depending on the assumed values of $p$ and $\Phi(\alpha)$.  Additional complications arise because published  observations use different filters,  and assume different phase functions and geometric albedos.   Albedos of fresh icy surfaces tend to be large, $p \sim$ 0.8 - 0.9, while the carbonaceous surfaces of primitive asteroids tend to be small, $p \sim$ 0.04. To proceed, we scale relevant properties assuming that $p$ = 0.1.  

With this albedo, the effective nuclear radius of `Oumuamua lies in the range 55 m \citep{Jewitt2017}, 70$\pm$3 m \citep{Meech2017}, 80  m \citep{Drahus18} to 114  m \citep{Knight2017}.  With no objective way to decide amongst these estimates, we take a middle value, $r_n =$ 80 m, as our best estimate of the effective radius, included in Table \ref{physical_properties}, and note that this value is uncertain by a factor of order two.

The bright coma  prevented direct detection of the nucleus of 2I/Borisov. However, \citet{Jewitt20} constrained the nuclear radius to lie within the range 0.2 $\le r_n \le$ 0.5 km. The upper limit is set by the non-detection of the nucleus in high resolution surface brightness data.  The lower limit is derived from the non-gravitational acceleration assuming a  comet-like bulk density $\rho_n$ = 500 kg m$^{-3}$.  We adopt $r_n$ = 400 m as the nominal value.

\subsection{Colors}
\label{colors}
Broadband colors are not compositionally diagnostic, but they do provide another metric with which to compare the interlopers with other solar system bodies. The conventional reference point is provided by the color of the Sun, which has optical color index B-R = 0.99$\pm$0.02.  For objects in which the albedo varies linearly with wavelength, as is commonly the case in distant solar system bodies, it is useful to use the reflectivity gradient, defined as the fractional change in the brightness per unit wavelength, $S'$ [\% (1000\AA)$^{-1}$], relative to that of the Sun.  The Sun, by definition, has $S'$ = 0 \% (1000\AA)$^{-1}$.  Objects with B-R $>$ 1.6, corresponding to $S' \sim$ 25\% (1000\AA)$^{-1}$, are described as ``ultrared'', and widely suspected to consist of irradiated, carbon-rich material \citep{Jewitt02}.

\begin{figure}
\begin{center}
       \includegraphics[scale=0.28,angle=0]{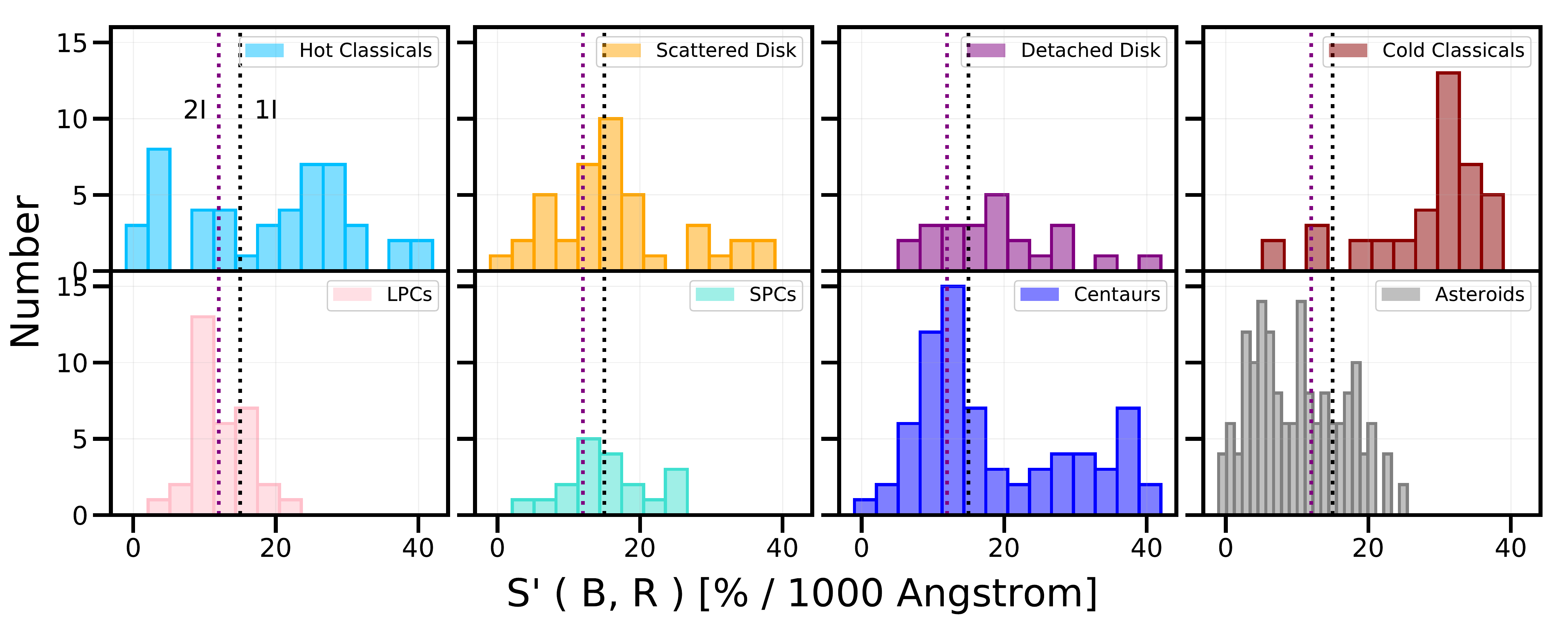}
    \caption{Color distributions of minor bodies in the solar system compared to those of `Oumuamua and 2I/Borisov (shown as vertical dotted lines).   
    Slope  $S'$ = 0 \% (1000\AA)$^{-1}$ corresponds to the solar color, B-R = 0.99, while $S' \sim$ 25 \% (1000\AA)$^{-1}$ corresponds to  B-R = 1.6. Colors are from \citet{Hainaut2012} and  \cite{warner2009database}.  } 
\label{fig:TNOColors}
\end{center}
\end{figure}

\textbf{1I/`Oumuamua:} 
  Published optical measurements of the color of `Oumuamua consistently  show a surface reddened with respect to the Sun, but with wide scatter.   This scatter reflects the difficulty of obtaining measurements for such a  highly variable, rapidly moving target.   The reflectivity  increases linearly with wavelength from at least 4500\AA~to 10,500\AA \citep{Fitzsimmons2017, Meech2017}. Values of the  spectral slope range from $S' =$ 7$\pm$3\%/1000\AA~\citep{Jewitt2017} to 23$\pm$3\%/1000\AA~\citep{Meech2017}; we adopt a middle value from independent measurements near $S' =$ 15$\pm$5\%/1000\AA.

\textbf{2I/Borisov:} Whereas the colors of 1I/`Oumuamua refer to the bare nucleus, those of 2I/Borisov measure the color of coma dust, with negligible contribution from the nucleus.  As with the case of `Oumuamua, reported measurements of the color of 2I show a reddened optical continuum with a scatter larger than the formal measurement errors \citep{Jewitt2019,Guzik2020,Bolin2020, Hui2020, Mazzotta21}.  Some of these differences in measurements may be attributed to the usage of a variety  of filters.  Specifically, it is known that the U and B filters  may be contaminated by resonance fluorescence bands from gas. The continuum color of 2I/Borisov  might also have varied with time and distance from the Sun \citep{Mazzotta21}.  We adopt a nominal spectral slope $S' \sim$ 12$\pm$1\% (1000\AA)$^{-1}$.

  In Figure \ref{fig:TNOColors}, we show the color distributions (expressed as $S'$) of different populations of minor bodies in the solar system. 
  These data are drawn from the online updated database\footnote{\href{http://www.eso.org/~ohainaut/MBOSS}{MBOSS Database: \url{https://www.eso.org/~ohainaut/MBOSS/}}} originally described in \citet{Hainaut2012}. We also indicate the positions of `Oumuamua and Borisov with respect to these distributions. 
  
  In the inner solar system, the reddish colors of asteroids are due to an abundance of nano-phase (1 to 100 nm) iron produced in surface materials by energetic particle bombardment (``space weathering''). Optical colors in the outer solar system can be much redder.  Ultrared colors  probably reflect the presence of irradiated, macro-molecular carbon compounds that are unstable or otherwise depleted in the warm, inner solar system.   The Cold Classical KBOs are the least dynamically evolved objects in the belt and located at about 43 AU. They also exhibit a larger fraction of ultra-red surface colors compared to most other small bodies in the solar system, with the exception of the Centaurs (recently escaped Kuiper belt objects).  This  difference is not understood, and may be attributed to longer exposure to space weathering if the Cold Classicals formed in situ.  The  colors  of 1I and 2I are similar to each other and  to most inner solar system populations (see the vertical dotted lines in Figure \ref{fig:TNOColors}).  The lack of distinctive evidence for  ultra-red matter, expected from long-term exposure to the interstellar environment, is consistent with the colors of LPCs.

       \begin{figure}
\begin{center}
\includegraphics[scale=.8,angle=0]{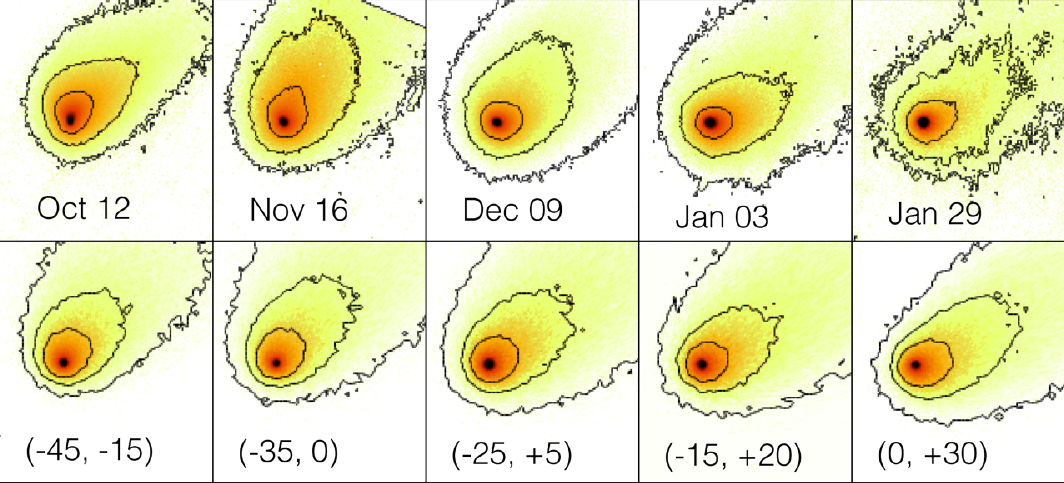}
    \caption{Images (upper row) and dust dynamics models (lower row) of 2I/Borisov as a function of date.   The models indicate a coma consisting of large (effective size $\sim$1 mm), slowly ejected ($V \sim$ 2 m s$^{-1}$) in steady-state.  Small particles are more strongly swept by radiation pressure, creating a coma and tail unlike those observed.  Adapted from \citep{Kim2020}. 
}\label{borisov_coma} 
\end{center}
\end{figure}

\subsection{Activity}

\textbf{1I/`Oumuamua:}
The morphology of `Oumuamua was at all times point-like, even in the highest resolution data from the Hubble Space Telescope and in the deepest data from large ground-based telescopes. The optical data provide no evidence for outgassing activity.   The optical scattering cross-sections of typical cometary comae are dominated by dust, with minor contributions from resonance fluorescence from gas molecules. 
 Accordingly, measurements of the surface brightness profile of `Oumuamua were used to place model-dependent limits on the mass loss in micron-sized dust particles; $\dot{M} \le 2\times 10^{-4}$ kg s$^{-1}$ \citep{Jewitt2017},  $\dot{M} \le 2\times 10^{-3}$ kg s$^{-1}$ \citep{Meech2017} (Table \ref{physical_properties}). Both upper limits are orders of magnitude smaller than the 10$^2$ kg s$^{-1}$ to 10$^3$ kg s$^{-1}$ mass loss rates from typical near-Sun comets estimated in the same way.  Limits on the production of water, the dominant cometary volatile, are $\dot{M} \le$ 30 kg s$^{-1}$, as determined from radio lines of the dissociation product, OH (see Section \ref{1Igas}).

\textbf{2I/Borisov:}
 In sharp contrast to `Oumuamua, 2I displayed (and, indeed, was discovered because of) obvious cometary activity in the form of an extended optical dust coma (Figure (\ref{1I_and_2I}) \citep{Jewitt2019,Guzik2020}.  Dust dynamics models reproduce the slowly changing morphology of the comet (Figure \ref{borisov_coma}) and show that the coma is dominated by sub-millimeter and larger particles \citep{Kim2020, yang2021}. Total dust production rates  estimated from imaging data alone range from $\dot{M} \sim$ 2 kg s$^{-1}$ \citep{Jewitt20} to  $\dot{M} \sim$ 35 kg  s$^{-1}$ \citep{Cremonese2020,Kim2020}. Less model-dependent spectroscopic production rates gave gas production rates up to $\dot{M}$ = 20 to 40 kg s$^{-1}$.

\begin{figure}
\begin{center}
\includegraphics[scale=0.27,angle=0]{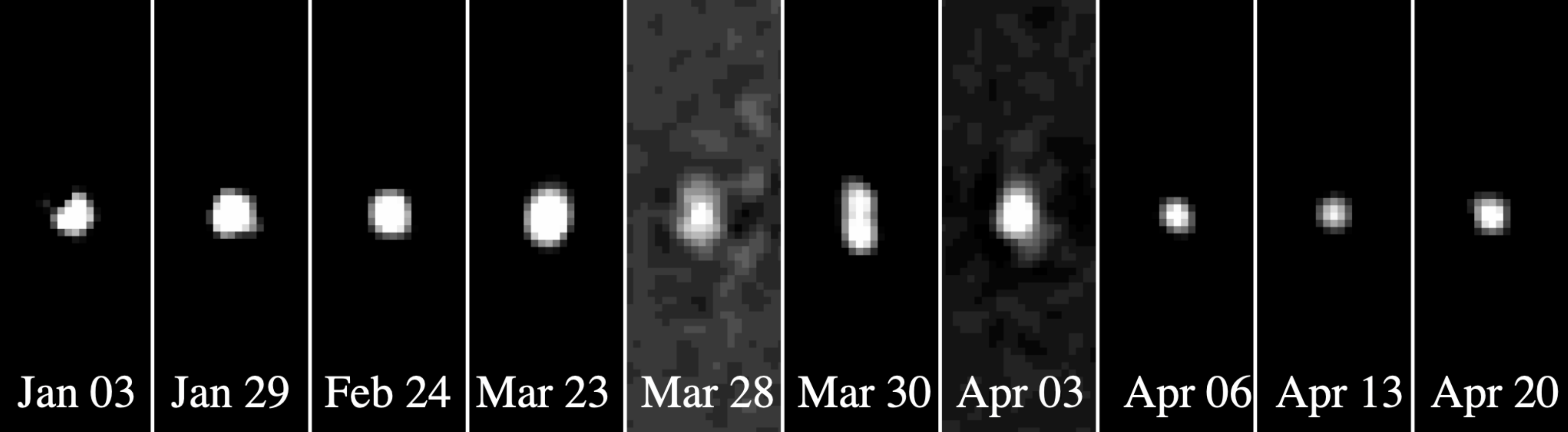}
    \caption{Spatially filtered (coma suppressed) Hubble Space Telescope images of 2I/Borisov showing the appearance of a split nucleus on UT 2019 March 30.  Each panel shows a region 0.44" wide corresponding to about 800 km at the comet. Note that the resolution of the telescope projects to $\sim150$km at the 2.65 AU geocentric distance of 2I/Borisov on March 30.  From \citep{Jewitt2020:BorisovBreakup}. }\label{borisov_split} 
\end{center}
\end{figure}

    Continued activity in 2I/Borisov included a photometric outburst \citep{Drahus2020ATel} and subsequent break-up, illustrated in Figure \ref{borisov_split} \citep{Jewitt2020:BorisovBreakup}.  The post-perihelion breakup might be related to a seasonal response, the likes of which are commonly observed in solar system comets \citep{Kim2020}.

  \subsection{Lightcurves}
\begin{figure}
\begin{center}
       \includegraphics[scale=0.23,angle=0]{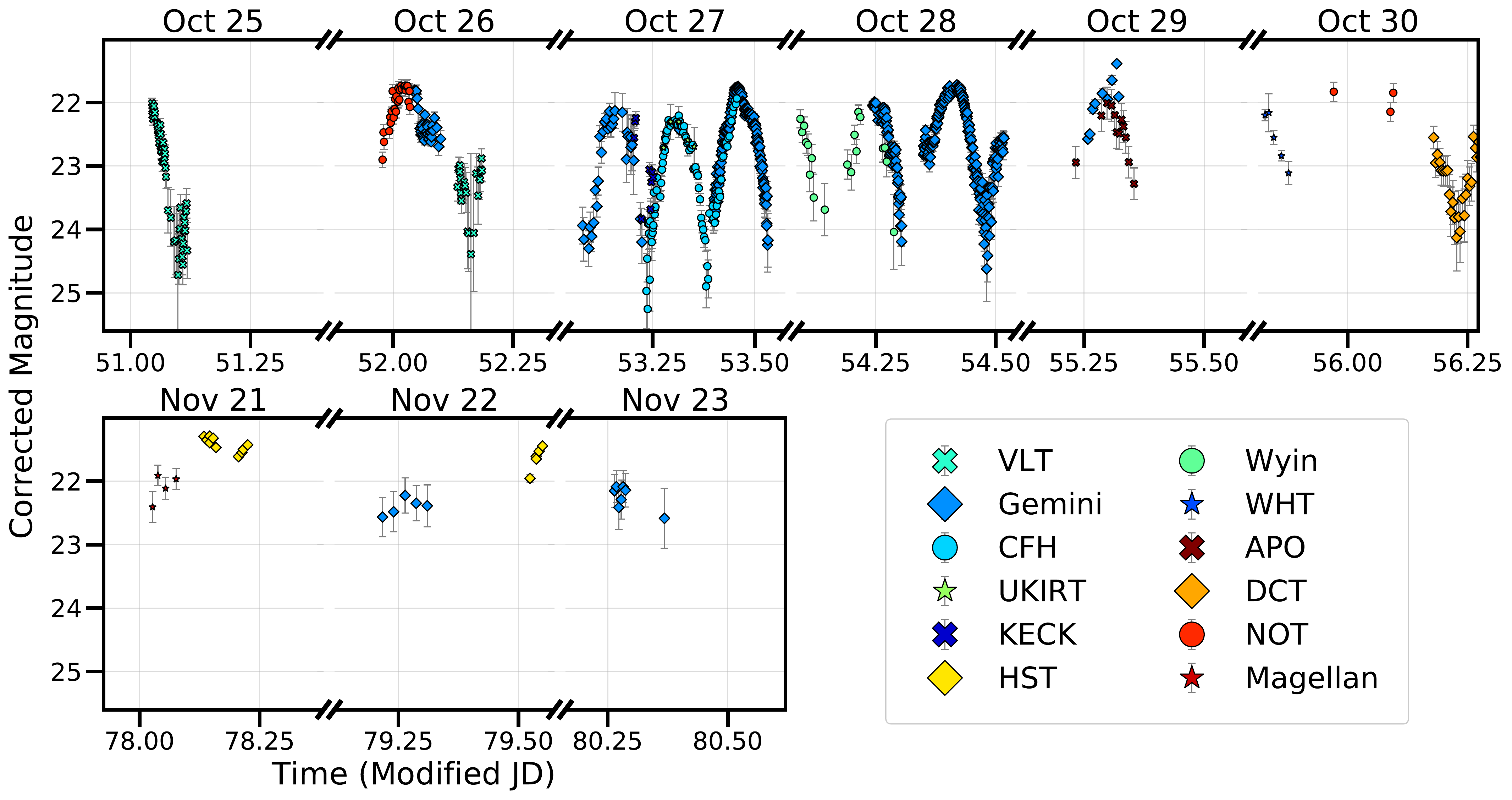}
    \caption{The photometric lightcurve of `Oumuamua, similar to that computed by \citet{Belton2018}. This figure uses the photometric data presented by \citet{Meech2017}, \citet{Jewitt2017}, \citet{Bannister2017}, \citet{Bolin2017}, \citet{Drahus18}, \citet{Knight2017} and \citet{Belton2018}. The points are color coded based on the observational facility that obtained the data.} 
\label{fig:light_curve} 
\end{center}
\end{figure}

\begin{figure}
\begin{center}
       \includegraphics[scale=0.27,angle=0]{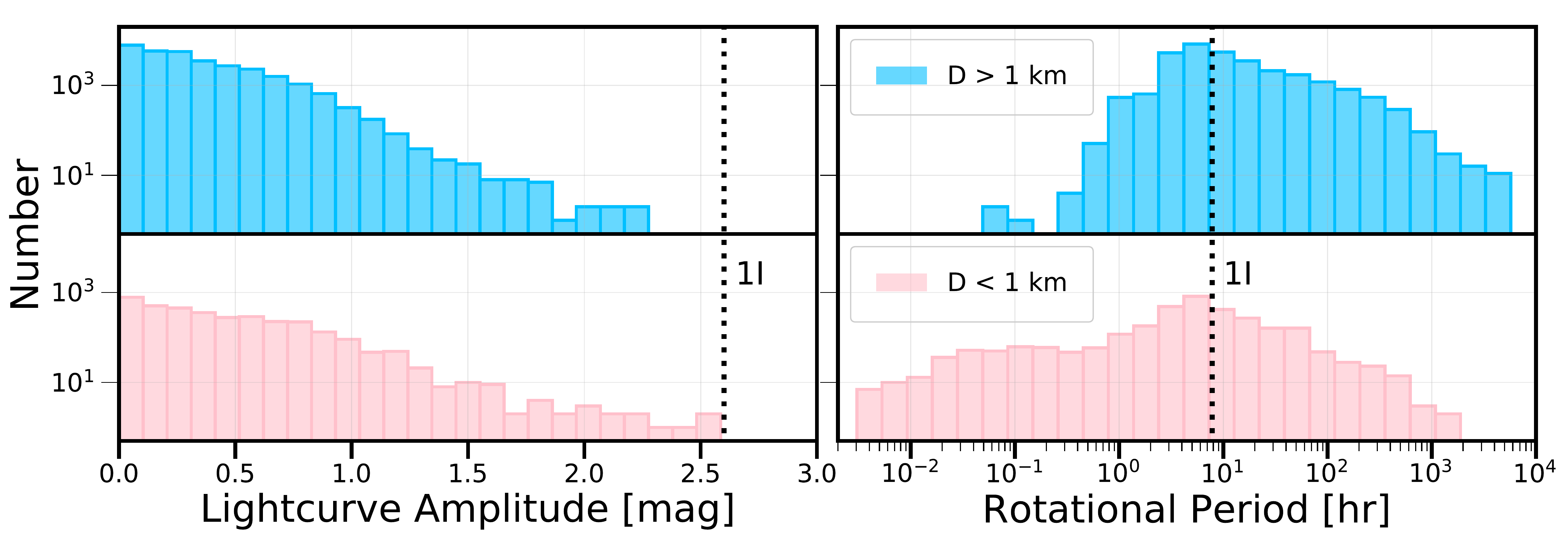}
    \caption{  The distribution of asteroid lightcurve amplitudes (left) and rotational periods (right).  The measured values for `Oumuamua are shown as dotted black lines (these are unconstrained for Borisov). The top and bottom panels show the distribution for all objects greater than and less than 1km in diameter, respectively. Data from  \cite{warner2009database}.  }
    \label{fig:amplitudes} 
\end{center}
\end{figure}

 \textbf{1I/`Oumuamua:} Photometric observations immediately showed that `Oumuamua had an extremely large lightcurve peak-to-peak amplitude of $\Delta V \sim$ 2.5 magnitudes, corresponding to a factor of $\sim$10 in brightness and a stable period close to 8 hours \citep{Meech2017,Jewitt2017,Knight2017,Bannister2017,Bolin2017,Drahus18}.  The lightcurve is shown in Figure \ref{fig:light_curve}, adapted from the data in Figure 1 of \citet{Belton2018}.     While the period of `Oumuamua is unremarkable compared with the distribution of rotational periods of the asteroids, the brightness variations  are extreme (Figure \ref{fig:amplitudes}), as will be discussed below.

 Lightcurves of atmosphereless solar system bodies are caused by azimuthal shapes or surface albedo variations  modulated by rotation.  To first order,  the  shape of an irregular body  projected into the sky-plane can be approximated as an ellipsoid with semi-axes $a \times b$. The ``effective radius'', equal to the radius of a circle having equal area,   is simply $r_n = (ab)^{1/2}$ (Table \ref{1I_and_2I}).  At opposition (Sun-target-Earth angle = 0$^o$), the axis ratio $b/a$ is related to the apparent lightcurve amplitude, $\Delta V$ in magnitudes, by   $b/a = 10^{0.4\Delta V}$. Substitution gives  $b/a$ = 10:1 for `Oumuamua \citep{Meech2017}. However, most  observations of `Oumuamua were obtained at phase angles $\alpha\sim$20$^o$, where illumination and self-shadowing effects act to magnify the lightcurve amplitude relative to that at zero phase \citep{Kaasalainen2001,Lacerda07, lu19}.  
 Taking these effects into account, the projected axis ratio of `Oumuamua corresponding to $\Delta V$ = 2.5 magnitudes is reduced to $b/a$ = 6:1 or 7:1 \citep{Jewitt2017,Drahus18,McNeill2018}.  This is still an extreme shape compared to known solar system bodies (c.f.~Figure \ref{fig:amplitudes}).   The shapes of  the asteroids are controlled by repeated energetic and sometimes  disruptive collisions. Even for these violent cases, the average projected axis ratios are $b/a \sim$1.4:1 and  values $b/a \ge$ 2 are  rare.   With $b/a$ = 6:1 and effective radius $r_n$ = 80 m, the projected sky-plane shape of `Oumuamua is an $a \times b$ = 32$\times$196 m ellipse. It is worth noting that the NEO 2016 AK$_{193}$ exhibited brightness variations of $\sim2.5-3$ magnitudes during its discovery apparition and subsequent followup observations \citep{Heinze2021DPS}.

Elongated solar system bodies are typically prolate in shape and rotating around a minor axis.  A large lightcurve amplitude can also be produced by an oblate body rotating around a long axis.  Detailed numerical and analytic calculations performed by \citet{Mashchenko2019} demonstrated that the most likely shape of `Oumuamua was a 6:6:1 oblate ellipsoid, as opposed to the 6:1:1 prolate geometry  popularized in several solicited artist impressions. Statistically, a randomly oriented oblate body with a given axis ratio is more likely than its prolate counterpart to yield a large rotational amplitude  consistently during each revolution. It is apparent in Figure \ref{fig:light_curve} that the object exhibited consistently deep brightness minima.  In this sense, we should assume that `Oumuamua is a flattened, disk-like body, not an elongated, cigar-shaped one.  Energy dissipation favors relaxation to the minimum energy (maximum moment of inertia) rotational state, which would work in favor of a prolate body shape, but the timescale for this relaxation depends on unknown physical parameters of the body and is suspected  to be very long \citep{Drahus18}.

  \textbf{Other Interpretations} It is possible that the lightcurve of `Oumuamua could result from azimuthal albedo variations instead of from projected shape variations.  In  solar system bodies, however, azimuthal albedo variations are almost always so small as to be immeasurable.  The physical reason is that the surface material is homogenized, both by space weathering and by gardening (the churning of exposed surface materials by  micrometeorite bombardment).  These processes, acting together or alone, render the surfaces of almost all asteroids and comets uniform in their scattering properties. The one notable exception to this is provided by Saturn's 1500 km diameter satellite Iapetus, which sports hemispheric $\sim$10:1  albedo variations.  Iapetus is a special case, however, because the satellite is in a spin orbit resonance, exposing one hemisphere to the impact of debris  from another Saturnian satellite while the other hemisphere survives unscathed \citep{Tamayo11}.   It is difficult to imagine how any comparable asymmetry could arise over the surface a body when free-floating in interstellar space.  Therefore, although the possibility that the lightcurve of `Oumuamua is caused by extreme azimuthal albedo variations cannot be formally rejected, it seems less plausible than an origin in the shape of the body.

  \textbf{2I/Borisov:} The optical cross-section of 2I/Borisov was dominated by dust in the coma, preventing the separation of the signal from the underlying nucleus.  For this reason, the rotational lightcurve could not be measured. Photometric variations in the coma instead provide a measure of the dust production rates in the object, subject to corrections for the changing viewing geometry.  2I/Borisov was observed over a wide range of phase angles, 40$^o \lesssim \alpha \lesssim$ 90$^o$, necessitating a large and correspondingly uncertain phase correction.  Depending on the details of this correction, some observers \citep{Jewitt2019} reported a net increase in dust cross-section on the approach to perihelion (commensurate with increasing gas production rates; Section \ref{1Igas})  while others inferred steady fading \citep{Hui2020}.  
   \subsection{Gas Production}
   \textbf{1I/`Oumuamua:}
   \label{1Igas}
No spectroscopic evidence for volatile outgassing  was found in `Oumuamua (\citet{Ye2017}, \citet{Fitzsimmons2017} and \citet{Trilling2018}).  We summarize the reported upper limits to production rates, $Q$, in Table \ref{Species}.  The corresponding mass production rates are $\dot{M} = \mu m_H Q$, where $\mu$ is the molecular weight of the species in question and $m_H$ is the mass of the hydrogen atom. From the Table we compute limits for water,  $\dot{M} \le$ 30 kg s$^{-1}$, and  for CO, $\dot{M} <$ 0.04 kg s$^{-1}$.

\begin{table}
\tabcolsep7.5pt
\caption{Upper limits to gas production rates in `Oumuamua.  }
\label{Species}
\begin{center}
\begin{tabular}{|@{}l|c|c|c|@{}l|}
\hline
Species  & Physical Property & Value& Distance& Reference\\
&& [molec s$^{-1}$ ]& AU&\\
\hline
{\rm CN } & Q({\rm CN})& $<2\times10^{22}$&1.4 AU &\citet{Ye2017}\\
{\rm C$_2$ } & Q({\rm C$_2$})& $<4\times10^{22}$&1.4 AU &\citet{Ye2017}\\
{\rm C$_3$ } & Q({\rm C$_3$})& $<2\times10^{21}$&1.4 AU &\citet{Ye2017}\\
{\rm OH } & Q({\rm OH})& $<1.7\times10^{27}$&1.8 AU &\citet{Park2018}\\
{\rm CO$_2$ } & Q({\rm CO$_2$})& $<9\times10^{22}$&2.0 AU &\citet{Trilling2018}\\
{\rm CO$^a$ } & Q({\rm CO})& $<9\times10^{23}$&2.0 AU &\citet{Seligman2021}\\
\hline
\end{tabular}
\end{center}
\begin{tabnote}
$^{\rm (a)}$ Corrected from Q({\rm CO})$<9\times10^{21}$ reported by \cite{Trilling2018} ]

\end{tabnote}
\end{table}

\begin{figure*}[t]
\begin{center}
       \includegraphics[scale=0.3,angle=0]{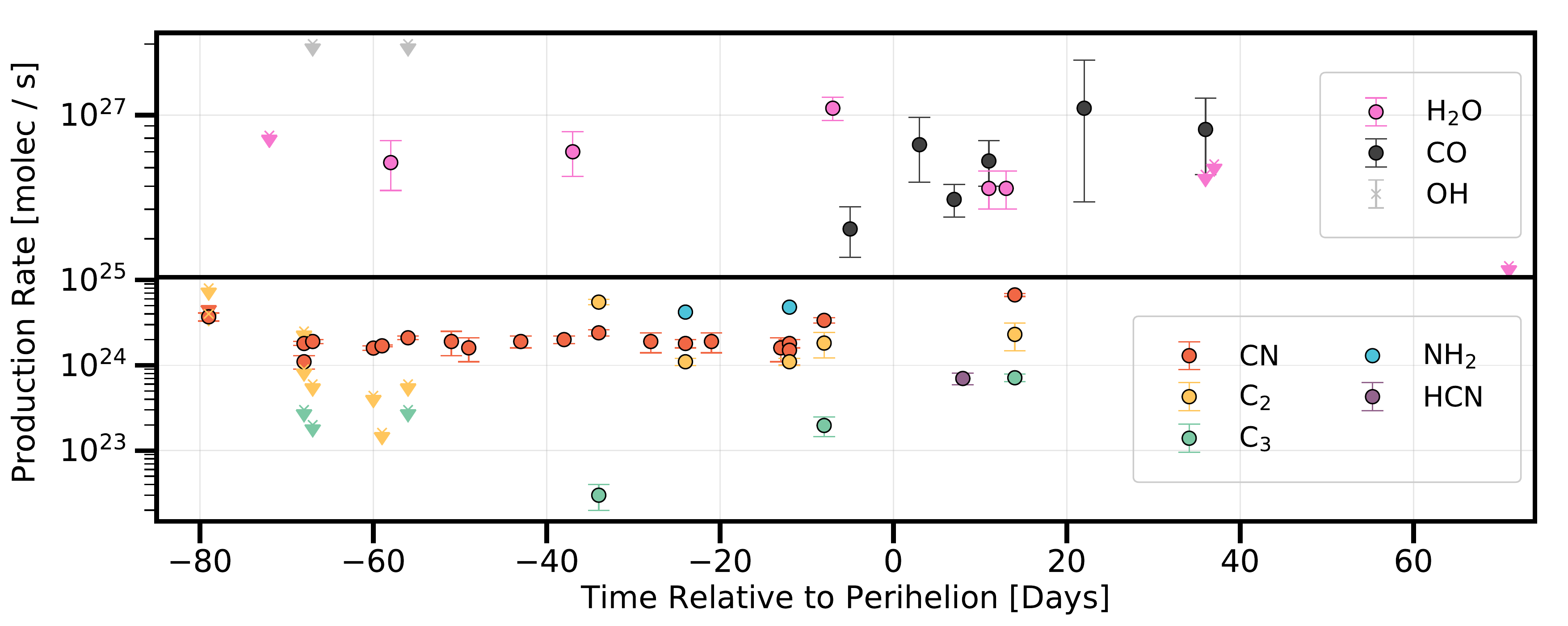}
    \caption{ 
    Time dependence of 2I/Borisov production rates. The top panel shows production rates of H$_2$O, CO and OH (Table \ref{Table:waterco}), and the bottom panel shows production rates of CN, C$_2$, C$_3$, NH$_2$ and HCN (Table \ref{Table:otherprods}). Adapted from \citet{Seligman2022PSJ}. }\label{fig:borisov_species} 
\end{center}
\end{figure*}

\textbf{2I/Borisov:}
    In contrast, the spectrum of 2I/Borisov showed distinct gas emission bands rising above its dusty continuum, corresponding to the classical resonance fluorescence features observed in short- and long-period comets.    Tables \ref{Table:waterco} and \ref{Table:otherprods} summarize the spectroscopic results.  The production rate data are also plotted in Figure \ref{fig:borisov_species} as a function of the time of observation.      The figure  shows a relatively small variation in the production rates, reflecting the small range of heliocentric distances over which data were obtained.

   \begin{table}[h]
   
\tabcolsep7.5pt
\caption{Production rates of CO, H$_2$O and OH measured for 2I/Borisov. Adapted from \citet{Seligman2022PSJ}.}
\label{Table:waterco}
\begin{center}
\begin{tabular}{@{}l|c|c|c|c|c@{}}
\hline
Date & r$_H^a$ [au]  & ${\rm Q({H_2O})}$ & ${\rm Q({CO})}$&  ${\rm Q({OH})}$ & Reference \\
&[AU]& $10^{26}$[s$^{-1}$]&$10^{26}$[s$^{-1}$]&$10^{26}$[s$^{-1}$]&\\
\hline
9/27/19&2.56&$<8.2$&&&\citet{Xing2020} \\\hline
10/2/19&2.50&&&$<0.2$&\citet{Opitom:2019-borisov}\\\hline
10/11/19&2.38&$6.3\pm1.5$&&&\citet{McKay2020} \\\hline
10/13/19&2.36&&&$<0.2$&\citet{Opitom:2019-borisov}\\\hline
11/1/19&2.17&$7.0\pm1.5$&&&\citet{Xing2020}\\\hline
12/1/19&2.01&$10.7\pm1.2$&&&\citet{Xing2020}\\\hline
12/3/19&2.01& &$3.3\pm0.8$&&\citet{yang2021}\\\hline
12/11/19&2.01&&$7.5\pm2.3$&&\citet{Bodewits2020}\\\hline
12/15-16/19&2.02&&$4.4\pm0.7$&&\citet{Cordiner2020}\\\hline
12/19-22/19&2.03 &$4.9\pm0.9$&$6.4\pm1.4$&&\citet{Bodewits2020}\\\hline
12/21/19&2.03 &$4.9\pm0.9$&&&\citet{Xing2020}\\\hline
12/30/19&2.07&&$10.7\pm6.4$&&\citet{Bodewits2020}\\\hline
1/13/20&2.16&$<5.6$&$8.7\pm3.1$&&\citet{Bodewits2020} \\\hline
1/14/20&2.17&$<6.2$&&&\citet{Xing2020}\\\hline
2/17/20&$2.54$&$<2.3$&&&\citet{Xing2020}\\\hline
\end{tabular}
\end{center}
\begin{tabnote}
$^{\rm a}$ Heliocentric distance at observation.
\end{tabnote}
\end{table}

      \begin{table}[h]
\tabcolsep7.5pt
\caption{Production rates of CN, C$_2$ and C$_3$ measured for 2I/Borisov. Adapted from \citet{Seligman2022PSJ}. }
\label{Table:otherprods}
\begin{center}
\begin{tabular}{@{}l|c|c|c|c|c@{}}
\hline
Date & r$_H^a$ [au]  & ${\rm Q({CN})}$ & ${\rm Q({C_2})}$ & ${\rm Q({C_3})}$ &  Reference \\
&[AU]& $10^{24}$[s$^{-1}$]&$10^{24}$[s$^{-1}$]&$10^{24}$[s$^{-1}$]&\\
\hline

9/20/19&2.67&$3.7\pm0.4$&$<4$&&\citet{Fitzsimmons:2019}\\\hline
9/20/19&2.67&$<5$&$<8$&&\citet{Kareta:2019}\\\hline

10/1/19&2.50&$1.1\pm2.0$&$<2.5$&&\citet{Kareta:2019}\\\hline
10/1/19&2.51&$1.8\pm0.1$&$<0.9$&$<0.3$& \citet{Opitom:2019-borisov}\\\hline
10/2/19&2.50&$1.9\pm0.1$&$<0.6$&$<0.2$&\citet{Opitom:2019-borisov}\\\hline
10/9/19&2.41&$1.59\pm0.09$&$<0.44$&& \citet{Kareta:2019}\\\hline
10/10/19&2.39&$1.69\pm0.04$&$<0.162$&&\citet{Kareta:2019}\\\hline

10/13/19&2.36&$2.1\pm0.1$&$<0.6$&$<0.3$&\citet{Opitom:2019-borisov}\\\hline
10/18/19&2.31&$1.9\pm0.6$&&&\citet{Opitom:2019-borisov}\\\hline
10/20/19&2.29&$1.6\pm0.5$&&&\citet{Opitom:2019-borisov}\\\hline
10/26/19&2.23&$1.9\pm0.3$&&&\citet{Kareta:2019}\\\hline
10/31/19&2.18&$2.0\pm0.2$&&&\citet{Lin2020}\\\hline
11/4/19&2.15&$2.4\pm0.2$&$5.5\pm0.4$&$0.03\pm0.01$& \citet{Lin2020}\\\hline
11/10/19&2.12&$1.9\pm0.5$&&& \citet{Bannister2020}\\\hline
11/14/19&2.09&$1.8\pm0.2$&$1.1$&& \citet{Bannister2020}\\\hline
11/17/19&2.08&$1.9\pm0.5$&&&\citet{Bannister2020}\\\hline
11/25/19&2.04&$1.6\pm0.5$&&&\citet{Bannister2020}\\\hline
11/26/19&2.04&$1.8\pm0.2$&&& \citet{Bannister2020}\\\hline
11/26/19&2.04&$1.5\pm0.5$&$1.1$&& \citet{Bannister2020}\\\hline
11/30/19&2.01&$3.36\pm0.25$&$1.82\pm0.6$&$0.197\pm0.052$&\citet{Aravind2021}\\\hline
12/22/19&2.03&$6.68\pm0.27$&$2.3\pm0.82$&$0.714\pm0.074$& \citet{Aravind2021}\\\hline
\end{tabular}
\end{center}
\begin{tabnote}
$^{\rm a}$ Heliocentric distance at observation.
\end{tabnote}
\end{table}


The  mass-dominant cometary volatile  in long and short period comets, H$_2$O, was  measured systematically in 2I/Borisov by \citet{Xing2020}.  They obtained observations at six epochs  before and after perihelion with the Neil Gehrels Swift Observatory's Ultraviolet/Optical Telescope. These data revealed a water production rate peaking near $Q({\rm H_2O}) = (10.7\pm1.2)\times 10^{26}$  s$^{-1}$ at perihelion, corresponding to mass production rate  $\dot{M}({\rm H_2O}) =$ 32 kg s$^{-1}$.   Adopting a nominal equilibrium sublimation rate of  water ice at 2 AU equal to $f_s = 5\times10^{-5}$ kg m$^{-2}$ s$^{-1}$ (obtained from solution of the energy balance Equation \ref{equilibrium}), the required area of exposed and sublimating water ice  is $C = \dot{M}({\rm H_2O})/f_s \sim0.64$ km$^2$. This surface area corresponds to that of a circle with radius $r_n = (C/\pi)^{1/2} \sim $ 0.45 km.  This ``sublimation radius'' lies close to the nominal 0.4 km nucleus radius and to the upper limit (0.5 km)   obtained  by  \citet{Jewitt2020:BorisovBreakup} in an independent calculation. These two independent estimates both corroborate each other and imply that the surface of 2I/Borisov had a large active fraction, $f_A \sim$ 1.  This, in turn, matches measurements showing that short-period comet nuclei of size comparable to 2I/Borisov typically have $f_A \sim 1$ \citep{Jewitt2021}.

2I/Borisov was also a productive source of carbon monoxide, CO, with a production rate ratio  measured on multiple  dates to be $Q({\rm CO}) / Q({\rm H_2O}) =$ 0.7$\pm$0.3 \citep{Cordiner2020}, and 1.3 to 1.6 \citep{Bodewits2020}. This is  substantially higher than  mean values $Q({\rm CO}) / Q({\rm H_2O}) =$ 0.04 in solar system comets at distances $\lesssim$ 2.5 AU \citep{Bockelee17, McKay2019}. The high relative abundance of CO implies low temperature formation of 2I/Borisov in order to trap the CO, presumably in the outer regions of a protoplanetary disk.  It also implies low temperature storage since formation in order to retain the CO against sublimation (see section \ref{section:normal}).

\section{Physical Models}

A central puzzle is that 1I/`Oumuamua showed no visible coma in deep composite images, and yet had a non-gravitational acceleration of 30$\sigma$ significance. These two observations have prompted  theories, from the mundane to the fantastical, regarding the provenance of the object.  In comparison, the properties of 2I/Borisov were closer to those observed in numerous solar system comets.

\subsection{Recoil from outgassing}
\label{recoil}
 Non-gravitational accelerations in solar system comets are  caused by  recoil  in response to the sublimation of surface ice in the heat of the Sun, and the subsequent  anisotropic mass ejection.  The recoil force may be written  as $k_R \dot{M} V_{s}$, where $\dot{M}$ is the sublimation rate, $V_{s}$ is the speed of the ejected material, and $0 \le k_R \le 1$ is a dimensionless constant representing the degree of anisotropy of the outflow.   For perfectly isotropic flow, $k_R$ = 0, while for perfectly collimated outflow, $k_R$ = 1. Newton's law applied to a body of spherical-equivalent radius $r_n$ and bulk density $\rho_n$ then gives,
 
 \begin{equation}
     \dot{M} = \,\bigg(\,\frac{4\pi \rho_n r_n^3}{3}\,\bigg)\,\bigg(\,\frac{ \alpha_{ng}(1)}{k_R V_s}\,\bigg)\,,
     \label{emdot}
 \end{equation}
 
 \noindent for the mass loss rate needed to generate the acceleration, $\alpha_{ng}(1)$.  
 
  Measurements of comets show that sublimation proceeds mostly from the hot dayside of the  nucleus. Correspondingly,  the recoil force acts primarily in the radial direction away from the Sun (i.e.~along $A_1$), as can be seen in Table \ref{dynamical_properties}.  The magnitude of $k_R$ has only been determined with confidence  for the short-period comet 67P/Churyumov-Gerasimenko at $k_R$ = 0.5, and is otherwise not well known. We adopt this value here. The speed of sublimated gas is close to the thermal velocity of gas molecules at the sublimation temperature of ice.  For water ice at $r_H \lesssim$ 2 AU, this is $T \sim$ 200 K, giving $V_s \sim$ 500 m s$^{-1}$, a value we adopt throughout.   The average density of solar system cometary nuclei is $\rho_n \sim$ 500 kg m$^{-3}$ \citep{Groussin19},  which we also adopt here.  Substitution into Equation \ref{emdot}, with our nominal spherical-equivalent radius estimate for `Oumuamua  $r_n$ = 80 m, gives $\dot{M} \sim$ 24 kg s$^{-1}$.  For the larger nucleus of 2I/Borisov, with 200 $\le r_n \le$ 500 m, mass loss rates $\dot{M}$ = 400 to 6000 kg s$^{-1}$ (scaled to 1 AU) are needed to account for $\alpha_{ng}(1)$.  Can these rates be supplied by sublimation of cometary volatiles?

The equilibrium rate of sublimation of a volatile surface exposed to the Sun is given by solutions to the following equation,

\begin{equation}
\bigg(\,\frac{ (1-A)\,L_{\odot}}{4\pi r_H^2}\,\bigg)\, \cos(\theta) =\, \varepsilon \sigma T^4 + H f_s(T) + C(T)\,.
\label{equilibrium}
\end{equation}

\noindent Here, $L_{\odot}$ (W) is the luminosity of the Sun, $r_H$ (m) the heliocentric distance, $A$ and $\varepsilon$ are the Bond albedo and thermal emissivity of the surface, $\sigma$ (W m$^{-2}$ K$^{-4}$) is the Stefan-Boltzmann constant, $T$ (K) is the surface temperature, $H$ (J kg$^{-1}$) is the latent heat of sublimation, and $f_s(T)$ (kg m$^{-2}$ s$^{-1}$) is the specific sublimation rate. The angle $\theta$ is the angle between the surface normal and the direction to the Sun. The term on the left hand side represents the absorbed solar power. The terms on the right hand side account for the thermal radiation, bond breaking in  sublimation and  conduction into the interior, respectively.  The surface materials on solar system bodies tend to be porous and  have low thermal conductivity, justifying the neglect of the conduction term $C(T)$  in most cases.  The latent heats of water and CO ice are $H = 2.8\times10^6$ J kg$^{-1}$ and $H = 2\times10^5$ J kg$^{-1}$, respectively.  Equation \ref{equilibrium} cannot be solved alone, and the Clausius-Clepeyron equation is commonly used in addition to represent the temperature dependence of the sublimation phase boundary in $f_s(T)$.  Moreover, the equation must be solved for a surface divided into elements, each with its own angle, $\theta$, to the solar direction.  The averaged value of $\cos(\theta)$ is as follows: $\overline{\cos(\theta)}$ = 1 for a flat surface oriented normal to the Sun, $\overline{\cos(\theta)}$ = 1/2 for a spherical body sublimating only from the sun-facing hemisphere, and $\overline{\cos(\theta)}$ = 1/4 for a uniformly sublimating  sphere.  $\overline{\cos(\theta)}$ = 1 represents the highest possible temperature, corresponding to the noon-day equatorial Sun, while $\overline{\cos(\theta)}$ = 1/4 corresponds to the lowest possible temperature on a sublimating, isothermal sphere.  The optical properties $A$ and $\varepsilon$ are generally not well constrained, but  Equation \ref{equilibrium} is insensitive to both provided $A \ll$ 1 and $\varepsilon \gg$ 0. Here, we assume $A$ = 0 and $\varepsilon$ = 0.9.

We adopt a nominal  reference distance of $r_H$ = 1 AU and assume that sublimation occurs from the sun-facing hemisphere of a spherical body (i.e.~$\overline{\cos(\theta)}$ = 1/2). Under these assumptions, Equation \ref{equilibrium} gives $f_s$(H$_2$O) = 2.2$\times10^{-4}$ kg m$^{-2}$ s$^{-1}$ and  $f_s$(CO) = 2.3$\times10^{-3}$ kg m$^{-2}$ s$^{-1}$ for the more volatile substance. We emphasize that these values are strictly valid only for sublimation at the surface.  A volatile that is protected from direct heat by an overlying layer of less volatile material (as is likely to be the case for highly volatile CO, for example) will sublimate at a smaller but much more model-dependent rate.  Production rates of water in solar system comets varies approximately as $r_H^{-2}$ out to $r_H \sim$ 2 AU.  Production rates of CO vary similarly, but out to $r_H \sim$ 50 AU.  This is due to the exponential rise in the sublimation term in Equation \ref{equilibrium} which is only in competition with the radiation term with $\sim T^4$ dependence. The model mass loss rate, $\dot{M}$, is given by $\dot{M} = 2 \pi r_n^2 f_s f_A$, where $f_A$ is the fraction of the nucleus surface from which ice sublimates.  By substituting this into Equation \ref{emdot},  the expected non-gravitational acceleration, $\alpha_{ng}$ for each volatile may be calculated.
\begin{figure}
\begin{center}
       \includegraphics[scale=0.23,angle=0]{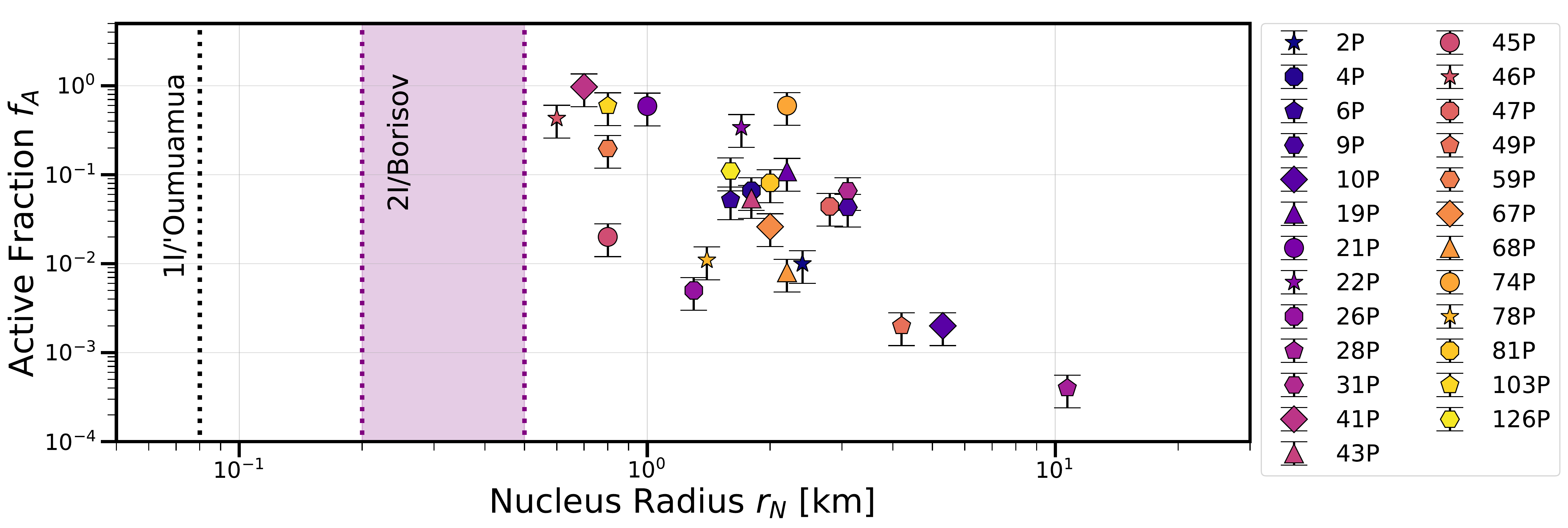}
    \caption{ Active sublimation surface fractions measured in solar system comets. Adapted from Figure 2 in \citet{Jewitt2021}. }\label{fig:fraction} 
\end{center}
\end{figure}
As shown in Figure \ref{fig:fraction}, the active fraction on short-period comets shows a clear trend with nucleus radius  \citep{Jewitt2021}, represented by a best-fit power law 
\begin{equation}
    f_A \simeq 0.1 \, r_N^{-2}\,,
\end{equation}
\noindent for $r_n \gtrsim$ 0.3 km (and $f_A$ = 1 otherwise). This trend is likely produced, at least in part, by observational bias, stemming from the fact that small nuclei with small active fractions are intrinsically faint and less likely to be discovered in flux-limited surveys. We plot and label the estimated radii of 1I and 2I in Figure \ref{fig:fraction}.  Both objects are small enough to suggest large active fractions, $f_A \sim$ 1, by analogy with the SPCs, and we adopt $f_A$ = 1 here. 

In the case of 1I/`Oumuamua, there are two immediate problems.  The first concerns the identity of the sublimating volatile.  Water ice sublimates too slowly at 1 AU to supply the $\dot{M} =$ 24 kg s$^{-1}$ needed to provide $\alpha_{ng}(1)$ as originally pointed out in an unrefereed preprint by \citet{Sekanina2019}.  Instead,  super-volatile ices (H$_2$, N$_2$, CO, Ne, Ar) sublimating in equilibrium with sunlight are required to generate sufficient recoil, as a result of their relatively small latent heats of sublimation. The noble gases have low abundance and are unlikely sources of activity. \citet{Seligman2020} proposed that 1I/`Oumuamua was composed of solid hydrogen, H$_2$, presumably formed in the failed  prestellar core of a giant molecular cloud (c.f.~\citet{Levine2021_h2}, \citet{hoang2020destruction}).  However, uncommonly frigid temperatures ($T <$ 6 K) are needed to accrete  and retain solid H$_2$ and the survival of such a volatile object in the open interstellar medium is in doubt. 

Next, \citet{desch20211i} and \citet{jackson20211i} proposed that sublimating N$_2$ could be the cause of the acceleration.  Nitrogen has the advantage of being spectroscopically inert, consistent with its non-detection in `Oumuamua. However, in the solar system, large nitrogen ice reservoirs are known to exist only on the surfaces of large, thermally differentiated Kuiper belt objects, with Pluto providing the premier example.  It is  unlikely that sufficient exposed solid nitrogen exists on differentiated extrasolar Kuiper belt objects to act as a galaxy-wide supply of 1I/`Oumuamua-like bodies \citep{Levine2021}.  
 
 Finally,  \citet{Seligman2021} suggested that carbon monoxide (CO) sublimation could supply the recoil acceleration. Equilibrium hemispheric sublimation of CO from an 80 m radius body at 1 AU is indeed sufficient to provide $\dot{M}$ = 24 kg s$^{-1}$ needed to accelerate `Oumuamua. Furthermore, CO is an abundant and observationally well-established volatile in solar system comets; its existence in `Oumuamua would not be surprising. Unfortunately, empirical limits to CO production in `Oumuamua \citep{Trilling2018} (even when corrected  for a numerical error to $Q_{CO} = 9\times10^{23}$ s$^{-1}$ ($\sim$0.04 kg s$^{-1}$) by \cite{Seligman2021}; Table \ref{Table:waterco}), are three orders of magnitude too small for CO to account for the non-gravitational acceleration.  The identity of a possible gaseous driver of the non-gravitational acceleration thus remains unresolved.

The second problem with  outgassing as an explanation of the non-gravitational acceleration of `Oumuamua is that no dust coma was detected.   Gas and dust mass production rates are comparable in typical comets, but the ejected dust typically dominates the optical appearance. This is because the dust has a much larger scattering cross-section per unit mass of material than does the gas, which is only rendered visible through resonance fluorescence.    Therefore, if sublimation caused the non-gravitational acceleration of 1I/`Oumuamua, it is particularly puzzling that no comet-like, dusty coma was evident even in the deepest images.  The empirical limits on the production rate of micron-sized particles ($<$(0.2 to 2)$\times 10^{-3}$ kg s$^{-1}$ (\citep{Jewitt2017,Meech2017}) are orders of magnitude smaller than required to supply the non-gravitational acceleration. 

A possible solution to this problem is that dust could be hidden from view given a sufficiently large  effective particle radius \citep{Micheli2018}.  The cross-section per unit mass of a collection of spheres with radius, $a$, varies as $\propto 1/a$.  Millimeter-sized particles in `Oumuamua would present 10$^{-3}$ of the cross-section of an equal mass of micron-sized particles. Consequently, they would be 10$^3$ times fainter in scattered light, perhaps allowing them  to have escaped detection.  This suggestion is ad-hoc  in the specific case of `Oumuamua, but measurements of some weakly active comets (e.g.~\cite{Ishiguro07}) and active asteroids (\cite{Jewitt22_asteroid}) indeed indicate mean particle radii of 10$^2$ $\mu$m and greater.  The physical explanation for this is not well understood, but a possible cause is that inter-particle cohesion (which itself varies as $1/a$) prevents small particles from escaping into the coma under the action of weak gas drag, leaving only the large particles to be ejected \citep{Skorov12}.   

Another possibility is that 1I/'Oumuamua outgassed through a porous mantle having enough strength to resist the expulsion of dust particles.  Again, this explanation is both ad-hoc and untestable. As a result, a consistent explanation of the origin of `Oumuamua's non-gravitational acceleration by outgassing has not been reached, and alternative explanations resting on the action of radiation pressure have been proposed (Section \ref{radiation}).

The case of 2I/Borisov is more clear-cut.  We possess independent estimates of the nucleus radius (200 $\le r_n \le$ 500 m), the non-gravitational acceleration (Table \ref{dynamical_properties}), and  the mass loss rate in gas (Table \ref{Table:waterco}). The latter is $\sim$80 kg s$^{-1}$ at 2 AU or $\dot{M} \sim$  320 kg s$^{-1}$ when scaled to 1 AU by the inverse square law.  Using these values to solve Equation \ref{emdot} for the nucleus density then gives 100 $\le \rho_n \le$ 1600 kg m$^{-3}$, a plausible range that brackets the nominal $\rho_n$ = 500 kg m$^{-3}$ density of solar system comets  (\cite{Groussin19}).  Ultra-low densities like that posited for `Oumuamua (Section \ref{radiation}) are specifically excluded from the allowable range of solutions for 2I/Borisov. The coma morphology was consistent with a dust differential power law size distribution index -3.5 and the absence of small particles.  The effective mean particle size was $a \gtrsim$ 100 $\mu$m \citep{Jewitt2019,Kim2020, Hui2020}, suggesting the role of small particle sticking as posited for 1I/`Oumuamua.  

\subsection{Radiation pressure: Fractal Bodies} 
\label{radiation}

 Radiation pressure offers an entirely different interpretation of the non-gravitational acceleration of 1I/`Oumuamua.  At 1 AU, the radiative pressure is $F_{\odot}/c$, where $F_{\odot}$ = 1360 W m$^{-2}$ is the solar constant and $c = 3\times10^8$ m s$^{-1}$ is the speed of light.  The force on a spherical body of radius $r_n$ is then $\pi F_{\odot} r_n^2/c$ and, by Newton's law, the density required to account for the measured non-gravitational acceleration may be written as
  
  \begin{equation}
      \rho_n = \,\bigg(\,\frac{3}{4 r_n}\,\bigg)\,\bigg(\,\frac{ F_{\odot}}{ c  \alpha_{ng}(1)}\,\bigg)\,.
      \label{fracdense}
  \end{equation}
  
  \noindent By substitution of $\alpha_{ng}(1)$ for `Oumuamua, we find that $\rho_n \sim 0.01(100/r_n)$ kg m$^{-3}$. This is two orders of magnitude less dense than air and implies a highly porous structure.  For comparison, the least dense artificial solid is Aerographite with $\rho$ = 0.2 kg m$^{-3}$ \citep{Mecklenburg12}.  Built of a complex assemblage of thin carbon sheets and tubes, Aerographite is still an order of magnitude  denser than  required of `Oumuamua by Equation \ref{fracdense}.

  Could such a low density material form naturally? One suggestion is that `Oumuamua could have a fractal structure produced by  Ballistic Cluster-Cluster Aggregation (BCCA) in an ultra-low energy protoplanetary disk environment \citep{moro2019fractal}.  The mass of a fractal, $m_f$, varies with its size, $a_f$, as $m_f \propto a_f^D$, and the density scales as $\rho_f \propto a_f^{D-3}$, where $D$ is the fractal dimension. Therefore, smaller values of $D$ correspond to less dense, more open and even ``stringy'' structures enveloping large void spaces. 
  
  Fractal aggregates have been investigated in the context of planetary accretion, where they offer several attractive features. These include tight dynamical coupling with the gas, providing a way to overcome the radial drift, fragmentation and bouncing barriers to the growth of large bodies \citep{Meakin88,Suyama08, Garcia20}.  Tiny (0.1 $\mu$m?) monomers would be well-coupled to the disk gas and therefore would collide gently, at speeds set initially by Brownian motion.   In BCCA, comparably sized clusters collide and stick, producing larger clusters of lower density.  The fractal dimension in BCCA is $D \sim$ 2, meaning that  the density varies as $\rho_f \propto a_f^{-1}$.  As the aggregate sizes grow to  $a_f \sim$ 1 to 10 mm, numerical models show that the density reaches a minimum value $\rho_f \sim 10^{-2}$ kg m$^{-3}$ (i.e.~comparable to that required of 1I/`Oumuamua by Equation \ref{fracdense}) \citep{Kataoka13}. At larger sizes, the particle clusters are compressed, first by about an order of magnitude due to ram pressure from the surrounding gas and then to comet-like densities at size scales $a_f \gtrsim10^2$ m,  by gravitational self-compression. The particles in a fractal structure are held together by incredibly weak van der Waals forces but, counter-intuitively, even with densities as small as $\sim10^{-2}$ kg m$^{-3}$, fractal structures could  survive the stresses induced by rotation and Solar tides  \citep{Flekkoy19}. Indeed, Aerographite has a tensile strength $\sim 10^3$ N m$^{-2}$ \citep{Mecklenburg12}.  
  
 There is some evidence for fractal structure in comets, but only on very small spatial scales. Sub-millimeter aggregates  gently collected from the coma of 67P/Churyumov-Gerasimenko  had densities $<$1 kg m$^{-3}$ \citep{Fulle15} and fractal dimensions $D$ = 1.7$\pm$0.1 \citep{Mannel16}. However, this density is still larger, by a factor of 10$^2$, than the 0.01 kg m$^{-3}$ density implied by Equation \ref{fracdense} for `Oumuamua.  Moreover, although some individual submillimeter particles in the  nucleus of 67P show low densities and fractal structure, a majority do not. The bulk density of the nucleus is a much more compact $\rho_n \sim$ 500 kg m$^{-3}$ \citep{Groussin19}.

 There are no known ultra-low density bodies of significant size in the solar system, but this could be a selection effect; even if the solar system formed with an abundance of gently-agglomerated, ultra low density bodies,  it is unlikely that any would survive today.  Compression or destruction by impact in later, more energetic phases of planetary accretion would erase any evidence of their existence.  Only fractal bodies ejected early to the collisionless environment of interstellar space would have had a chance to survive.
 
 As a novel caveat to this idea, \cite{Luu20} suggested that fractal bodies might actively form in the comae of other comets, as particles lifted from the surface collide and stick to form  ``dust bunny'' aggregates.   While most cometary fragments appear short-lived, in this scenario we should be able to observe cometary fragments that are progressively accelerated into hyperbolic orbits.

\begin{table}
\tabcolsep7.5pt
\caption{Empirical spin-changes of short-period comet nuclei   }
\label{Table:comet_spinup}
\begin{center}
\begin{tabular}{@{}l|c|c|c|c|c@{}}
\hline
Comet Name  & q$^{\rm a}$ & $P_K^{\rm b}$& P$^{\rm c}$& $\Delta P^{\rm d}$& Reference\\
&[AU]& [year ]& [hour]&[minutes]\\
\hline
2P/Encke & 0.337& 3.30&11.0 &4&\citet{Roth2018}\\
\hline
9P/Tempel & 1.542& 5.58&40.9 &13.5&\citet{Gicquel2012}\\
\hline
10P/Tempel & 1.423& 5.37&8.9 &0.27&\citet{Wilson2017}\\
\hline
14P/Wolf & 2.729& 8.80&9.0 &$<4.2$&\citet{Fernandez2013}\\
\hline
19P/Borrelly & 1.358& 6.86&29.0 &20&\citet{Maquet2012}\\
\hline
41P/TGK & 1.046& 5.42&34.8 &1560&\citet{Combi2020}\\
\hline
46P/Wirtanen &1.055& 5.44&9.15 &12&\citet{Combi2020}\\
&&&&&\citet{Farnham2021}\\
\hline
49P/Arend–Rigaux &1.343& 6.62&13.0 &$<0.23$&\citet{Eisner2017}\\
\hline
67P/C-G &1.244& 6.45&12.0 &21&\citet{Biver2019}\\
\hline
103P/Hartley &1.058& 6.46&18.2 &120&\citet{Drahus2011}\\
&&&&&\citet{Combi2020}\\
\hline
143P/Kowal–Mrkos &2.542&8.90&17.0 &$<6.6$&\citet{Jewitt2003b}\\
\hline
162P/Siding Spring &1.232&5.30&33.0 &$<25$&\citet{Fernandez2013}\\
\hline
\end{tabular}
\end{center}
\begin{tabnote}
$^{\rm a}$ Perihelion; $^{\rm b}$Keplerian orbital period; $^{\rm c}$Nucleus spin period ; $^{\rm d}$measured change in nuclear spin period \textit{over the course of 1 orbit} \citep{Kokotanekova2018}. Adapted from \citep{Jewitt2021}.  .
\end{tabnote}
\end{table}

\subsection{Radiation pressure: Membrane}
\label{section:alien}

The measured acceleration of `Oumuamua, if due to radiation pressure, could instead imply a thin sheet geometry with a low column density, $\Sigma$ (kg m$^{-2}$), given by
  
  \begin{equation}
      \Sigma =\,\bigg(\, \frac{F_{\odot}}{c \,\alpha_{ng}(1)}\,\bigg)\,.
  \end{equation}
  
  \noindent Substitution for the case of `Oumuamua gives $\Sigma \sim$ 0.8 kg m$^{-2}$, a value that is more typical of a thin sheet of cardboard (density $\sim10^3$ kg m$^{-3}$, thickness $\sim10^{-3}$ m) than of any natural, macroscopic object.

\citet{Bialy18} suggested that `Oumuamua could be a razor-thin sheet or membrane, perhaps akin to a light-sail.  Such a structure would have a column density small enough to be accelerated by radiation pressure.  With a dimension $r_n \sim$ 100 m, the corresponding sheet mass would be $r_n^2 \Sigma \, \sim$ 10$^4$ kg, or smaller if the albedo is higher. Because the orbit of `Oumuamua is gravitationally unbound, they inferred that `Oumuamua could be the manufactured product of an alien civilization. Active radio signals transmitted from `Oumuamua were not detected \citep{Enriquez2018,Tingay2018,Harp2019}.

The alien membrane hypothesis is consistent with the  existence of non-gravitational acceleration without detectable mass loss, provided the membrane maintains an orientation nearly perpendicular to sunlight. It is also qualitatively consistent with `Oumuamua's extreme lightcurve, albeit with a low ($\sim$1\%) probability for having the  orientation needed to generate the observed large amplitude \citep{Zhou2022}. The alien membrane hypothesis is highly questionable on other grounds, however.  For example, `Oumuamua  cannot be a  probe targeted at Earth because it  missed the Earth by $\sim$40 million km; intelligent aliens could surely do better.  Could `Oumuamua  instead be a piece of alien space trash?  If so, it is difficult to see why  an intelligent civilization would flood the galaxy with $10^{25}$ or $10^{26}$ (Section \ref{sec:limits}) pieces of 100 meter scale, Mylar-like debris.

\subsection{Stability against spin-up destruction}

Models of the unexpected properties of 1I/'Oumuamua thus require either strained explanations in terms of the outgassing of unseen volatiles, or explanations invoking material of such low column density that radiation pressure has a strong effect.  Whichever model applies, the strong non-gravitational acceleration of this object implies that substantial torques should have modified the spin of the nucleus, potentially driving it to rotational instability \citep{Rafikov2018b}.

\begin{figure}
\begin{center}
       \includegraphics[scale=0.23,angle=0]{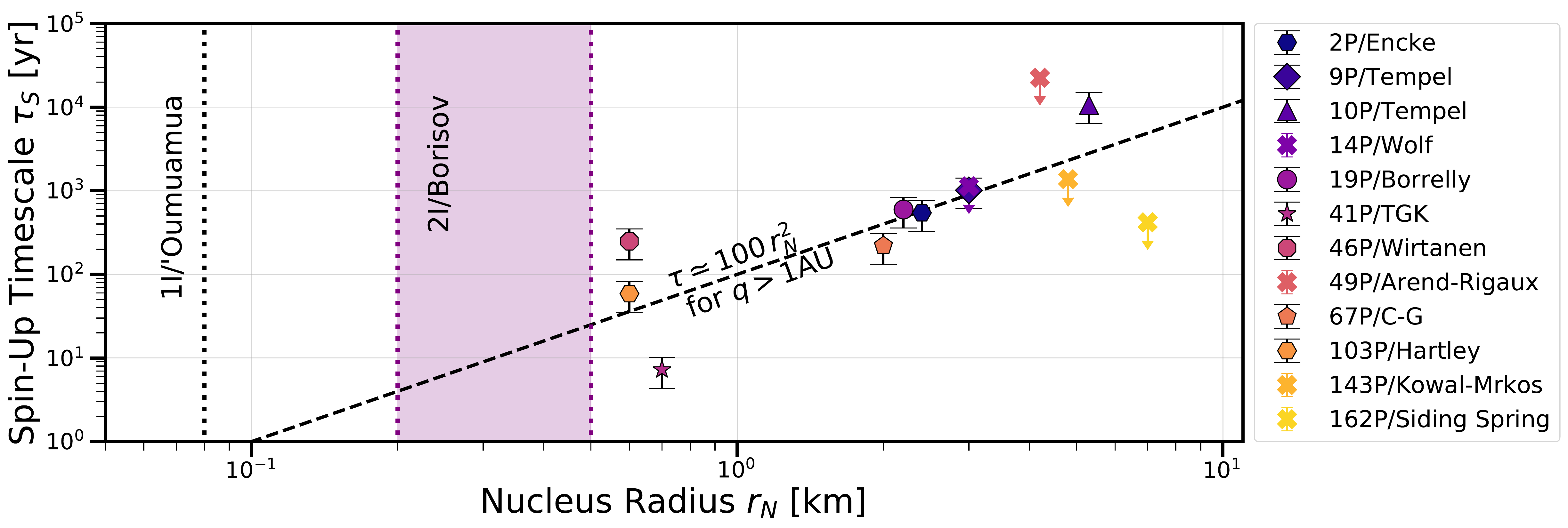}
    \caption{ Spin-up timescales measured in solar system comets. 
    Equation \ref{tau_s} is shown as a dashed line. Adapted from Figure 1 in \citet{Jewitt2021}. }\label{fig:spinup} 
\end{center}
\end{figure}

The nuclei of short-period comets show clear evidence for the action of outgassing torques, in the form of small changes in the rotation period per orbit, $\Delta P$ \citep{Kokotanekova2018}.  When sustained over sufficiently long times, these outgassing torques can drive comets to rotational instability, in which centripetal forces exceed those of gravity and material cohesion acting to bind the nucleus.  The result is nucleus breakup or disintegration, widely observed in both  short-period \citep{Jewitt2021} and long-period \citep{Jewitt22} comets.

Empirically, the timescale for changing the nucleus spin can be computed from 

\begin{equation}
    \tau_S=\,\bigg(\,\frac{P}{\Delta P}\, P_K\bigg)\,,
\end{equation}

\noindent where, $P$ is the measured rotational period, $\Delta P$ is the change in rotational period,  and $P_K$ is the Keplerian orbital period.  Table \ref{Table:comet_spinup} lists measurements of $\Delta P$ while Figure \ref{fig:spinup} shows $\tau_S$ as a function of nucleus radius for the nuclei of short-period comets having perihelion distance 1 $\le q \le$ 2 AU.  To a good level of approximation, the timescale is represented by 

\begin{equation}
    \tau_S \simeq 100\, r_N^{\,2} \textrm{~~(years)}\,,
    \label{tau_s}
\end{equation}

\noindent where $r_N$ is the nucleus radius expressed in km.  Equation \ref{tau_s} is shown in the figure as a dashed line.

The small nuclei of 1I/`Oumuamua and 2I/Borisov, both indicated in Figure \ref{fig:spinup},  have very short spin-up times, with that for `Oumuamua being $<$1 year. In fact,  1I/`Oumuamua should spin-up even more quickly than indicated by Equation \ref{tau_s}, both because its small perihelion distance would induce larger outgassing rates than in the comparison comet population, and because of its extreme aspect ratio. The latter would give a longer lever arm than for a more nearly spherical nucleus.  Based on these considerations, \citet{Rafikov2018} argued that the lack of steady spin-up and constant periodicity in the lightcurve implied that outgassing could not explain the anomalous acceleration.  

\begin{figure}
\begin{center}
       \includegraphics[scale=0.35,angle=0]{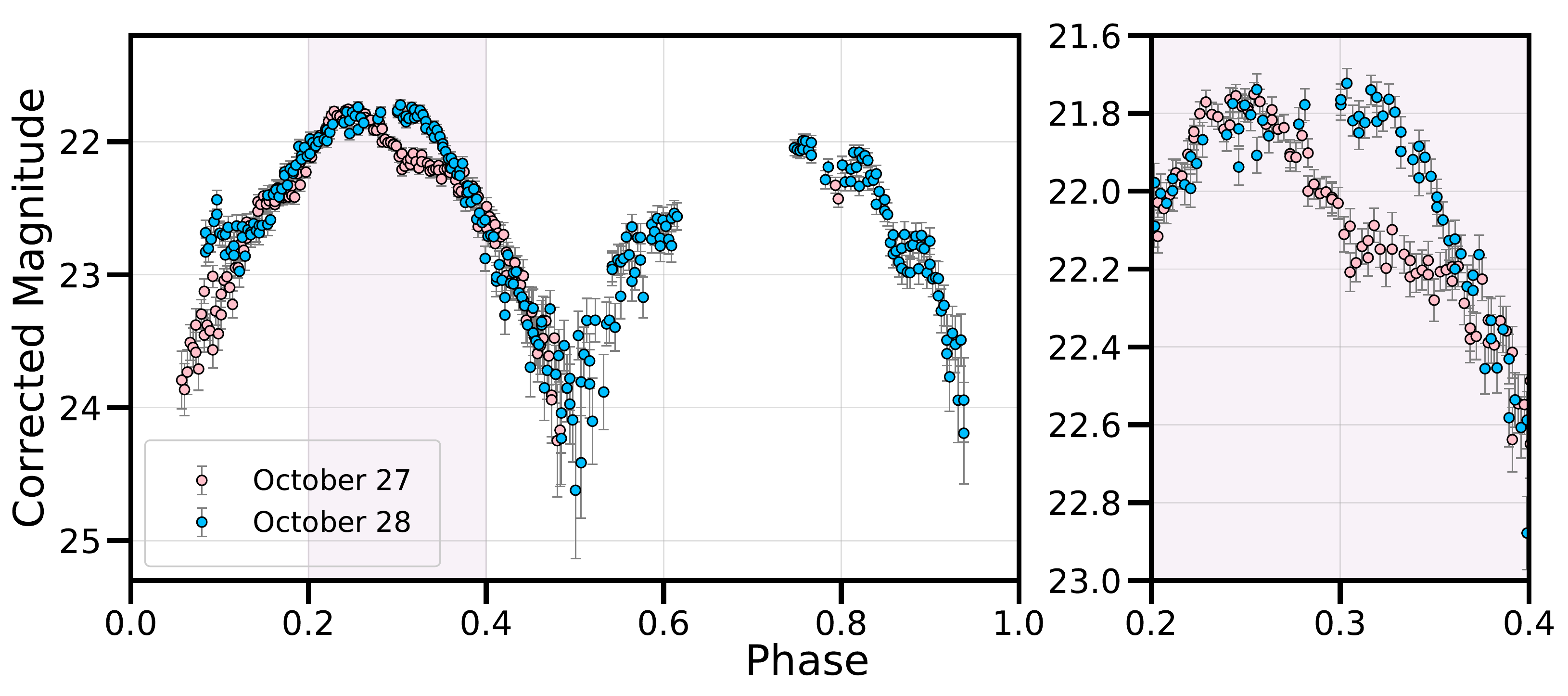}
    \caption{(left panel:) Phased lightcurve data from 1I/`Oumuamua showing small differences between measurements three rotations apart, from UT 2017 October 27 (pink cicles) and October 28 (blue circles). The data are phase folded at a period of 7.5483 hr. (right panel:) Zoom view of the region near phase 0.3 to show differences in the brightness. Adapted from \citet{Drahus18}. }
 \label{fig:phased} 
\end{center}
\end{figure}

Figure \ref{fig:phased} shows the phased lightcurve of `Oumuamua with photometric data obtained from only one telescope  and presented by \cite{Drahus18}.  The lightcurve is highly -- but not completely -- repetitive. For example,  differences at the $\sim$10\% level  are evident near phase = 0.01 and 0.35.  If these differences are not due to measurement error, they could indicate a slightly excited or ``tumbling''  rotational state, which  \cite{Drahus18} interpreted  as the result of an ancient collision in the protoplanetary disk of another star.  Rotation can also be excited contemporaneously by mass loss and other torques.  

\citet{Belton2018, Drahus18}, and \cite{Fraser2017} searched for periodicities in the composite lightcurve, after correcting the data for the changing phase angle and distance to `Oumuamua and for the use of different filters by different observers.  An analysis by \cite{Belton2018} identified dominant periodic signals at $8.67\pm0.34$ and $3.74 \pm 0.11$ hrs, consistent with a single rigidly rotating and precessing nucleus and these authors found no evidence for the action of torques on the rotation of 1I/`Oumuamua.

However, the \cite{Belton2018} analysis assumes a simple decomposition into discrete frequencies, potentially concealing evidence for the action of torques.  \cite{Flekkoy19} noticed that reported rotation periods of 1I/`Oumuamua increased linearly from 7.3 hour to 8.2 hour over the three day period UT 2017 October 26-29, corresponding to a spin-change timescale $\tau_s = P/\Delta P \sim$ 26 days. This is broadly consistent both with the YORP spin-up timescale for an ultra-low density body only 80 m in radius and with the outgassing torque timescale computed from Equation \ref{tau_s}.  An independent analysis revealed that the addition of a torque significantly improves the fit to the lightcurve \cite{Mashchenko2019}. Moreover, the best-fit moment arm for the torque ($k_T$ = 0.0046) is similar to the median moment arm measured in short-period comets ($k_T$ = 0.007; \citep{Jewitt2021}.  (The same study also revealed that an oblate body shape is more likely than a prolate one). 

Although different in detail, the \cite{Flekkoy19} and \cite{Mashchenko2019} studies still reveal fractional changes in the rotation period of order unity on timescales of $\sim$1 month, leaving it unclear how to reconcile the survival of the nucleus against rotational disruption with the large non-gravitational acceleration.   Point-source outgassing from the subsolar point would eliminate secular spin-up \citep{Seligman2019} but this pathological geometry does not occur on the known comets. Likewise, a symmetric elongated body with jets uniformly covering the  illuminated surface can produce stable spin dynamics \citep{Seligman2021} but, again, this would be unlike outgassing from any of the known comets. 

One intriguing possibility is that `Oumuamua \textit{did} rotationally disrupt from outgassing torques near perihelion, and that the body observed in 2017 is merely the remnant of a rotationally disrupted precursor nucleus that did not survive perihelion.  The drifting spin period  reported in Figure 1 of \cite{Flekkoy19}, when extrapolated backwards, reaches zero in early October, a few weeks after perihelion.  Moreover, this possibility is consistent with the observed disintegration of the nuclei of long-period comets, a likely result of rotational instability when near the Sun \citep{Jewitt22}.  Such rotational disintegration occurs preferentially in sub-kilometer nuclei with perihelia $<$1 AU; both conditions (radius $\sim$0.08 km, perihelion 0.25 AU) apply to `Oumuamua.  Residual outgassing from volatiles exposed by the breakup could further modify the rotation, and provide the  non-gravitational acceleration.

2I/Borisov, with its larger nucleus and perihelion distance, is less susceptible to rotational breakup, for which the timescale (Equation \ref{tau_s}) is 4 $\le \tau_s \le$ 25 years.  It did, however, release several fragments  (Figure \ref{borisov_split}), possibly under the influence of nucleus rotation.

\subsection{Tidal Remnant}  Comets in the solar system are occasionally tidally disrupted when passing within the Roche lobe of the Sun or a giant planet.  The most famous and best-studied example is that of the kilometer-sized, periodic comet P/Shoemaker-Levy 9 (1993 F2, ``SL9''), which disrupted upon passing Jupiter at 1.6 times the planetary radius in 1992 \citep{Weaver95}.  The example provided by SL9 motivated \citet{Raymond18} to consider `Oumuamua as a tidally shredded fragment of a precursor cometary body from the protoplanetary disk of another star.  They obtained an estimate of the fractional disruption rate of ejected bodies in the range 10$^{-3}$ to 10$^{-2}$, although this number sensitively depends on the model assumptions. They also pointed out that a majority of disrupted fragments would be de-volatilized by subsequent close approaches to the host star prior to escape (further investigated in \citet{Raymond2020}).  The fragment shapes produced by tidal disruption, subject to assumptions about the stellar impact parameter, rubble-pile structure, internal friction and more, are prolate and can be as elongated as suggested by the lightcurve of `Oumuamua  \citep{Zhang20}.  However, the galactic rate at which cometary disruptions and ejections occur, given the many unknowns, is highly uncertain.

Other potential sources of tidal fragments exist including post main-sequence stars \citet{Hansen2017,Rafikov2018b,Katz2018}, close stellar flybys in clusters \citep{Pfalzner21} and circumbinary systems \citep{Cuk2017,Jackson2017}. \citet{Childs2022} argued that \textit{misaligned} circumbinary disks  are particularly efficient progenitors of interstellar asteroids, specifically of objects that are close enough to  have lost their volatiles. 

In addition to being tidally shredded, an object must be ejected from the gravitational control of its host star if it is to join the interstellar rank.  In a given planetary system, the ejection efficiency depends on  the stellar mass, $M_{\star}$, and on the distance, $a_P$, size, $R_P$, and mass, $M_P$, of the scattering planet.  The distance and mass of the star set the Kepler velocity, $V_{K} = (GM_{\star}/a_P)^{1/2}$. The size and mass of the planet set the escape velocity from the surface of planet, $V_e = (2GM_P/R_P)^{1/2}$. The escape velocity is a reasonable approximation for the maximum velocity that can be imparted in a scattering event.

\begin{figure}
\begin{center}
       \includegraphics[scale=0.35,angle=0]{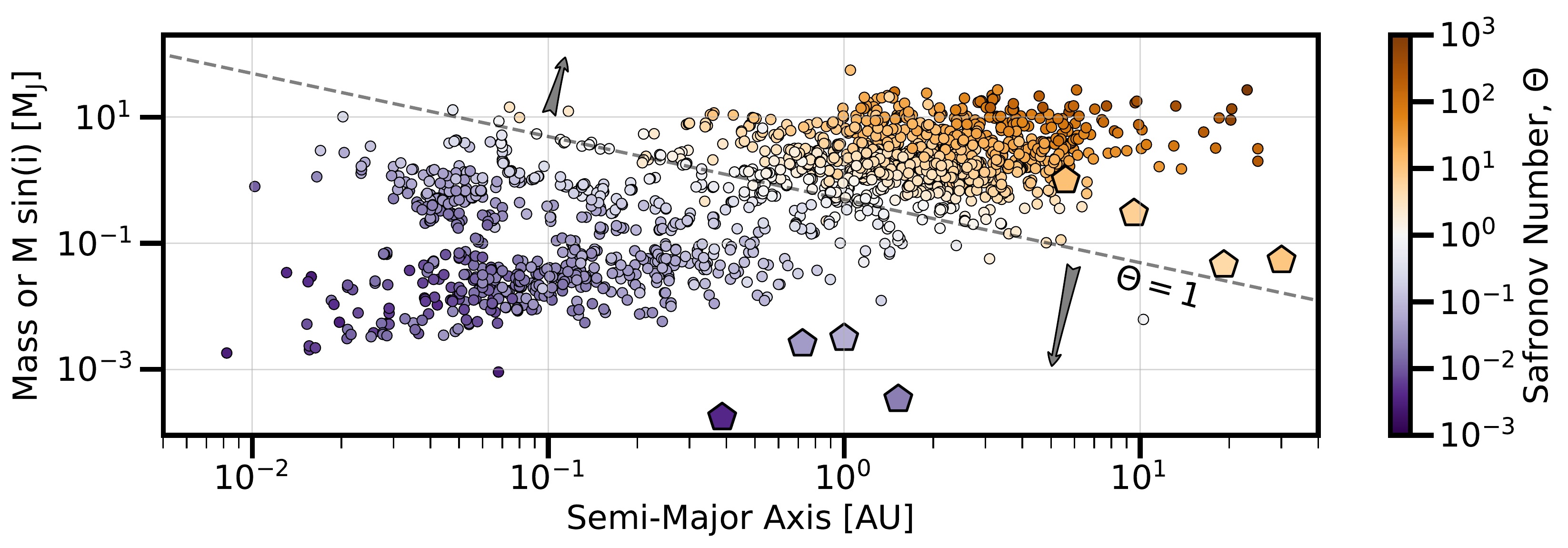}
    \caption{Safronov number (color coded) for the known exoplanets showing that those with $\Theta >$ 1 are preferentially located in the outer regions of their systems (c.f.~\citet{Laughlin2017}).  Large pentagonal symbols show solar system planets. The illustrative dashed line shows $\Theta=1$, for a Jupiter radius planet orbiting a solar mass star. For the most part, planets above the line are capable of ejecting comets to the interstellar medium while those below it are not. }\label{fig:safronov} 
\end{center}
\end{figure}

A useful parameter to  quantify the efficiency of ejection for any given perturber is the Safronov number, $\Theta = V_e^2/(2V_K^2)$ or, equivalently,

\begin{equation}
  \Theta= \left(\frac{a_{\rm P}}{R_{\rm P}}\right)\left(\frac{M_{\rm P}}{M_*}\right) \,.
\end{equation}

\noindent To a very good approximation, only planets with $\Theta>1$ can eject objects via scattering. In the modern-day solar system, the terrestrial planets  have $\Theta <$ 1 but all four giant planets satisfy $\Theta >$ 1 and are capable of ejecting comets into the interstellar medium.  The mass of material that can be ejected depends on the details of the planet-disk interaction. Importantly, for all planets where $\Theta>1$, larger values do not necessarily imply that the planet ejected more debris.  For instance, Jupiter has a much higher mass than Neptune, and naively one would imagine that it dominated the cometary ejection in the Solar System. However, Neptune migrated over a larger range of distances than did  Jupiter, providing it with access to a larger mass of nearby cometesimals. To complicate matters further, the density of the disk near Neptune was smaller than at Jupiter.  \cite{Hahn99} modeled  this process and found that the Oort cloud emplacement efficiency is broadly distributed across the Jupiter to Neptune region.

In Figure \ref{fig:safronov}, we show the Safronov number for currently confirmed extrasolar planets (c.f.~Figure 1 in \citet{Laughlin2017}). The figure shows that ejection through planetary scattering is unlikely to occur from the inner regions of the known planetary systems, consistent with the idea that most interstellar interlopers are ice-rich bodies formed beyond their snow-lines (see the discussion in Section \ref{section:normal}). Unfortunately, given the uncertainties in the architectures and evolution of other planetary system, it is not possible to meaningfully estimate the galactic rate of ejection to the interstellar medium other than by measurements of the interloper population. 

Whatever the mechanism, objects that are ejected from close to the host star must acquire higher velocities than those launched from more distant locations in order to overcome the gravity of the star. These higher velocities require scattering from more massive planets and, in planetary systems like our own, high velocity ejections should be rare.  Therefore, unless the progenitor system is a special case like a circumbinary system where it is easier to eject from closer in, ejection at high velocity  should be less common. A first order corollary of this is that the distribution of velocities of interstellar objects should closely resemble that of the stars upon ejection\footnote{These dispersions would be modified by subsequent dynamical heating.}.

\subsection{2I as a more "normal" comet}
\label{section:normal}

\begin{figure}
\begin{center}
       \includegraphics[scale=0.23,angle=0]{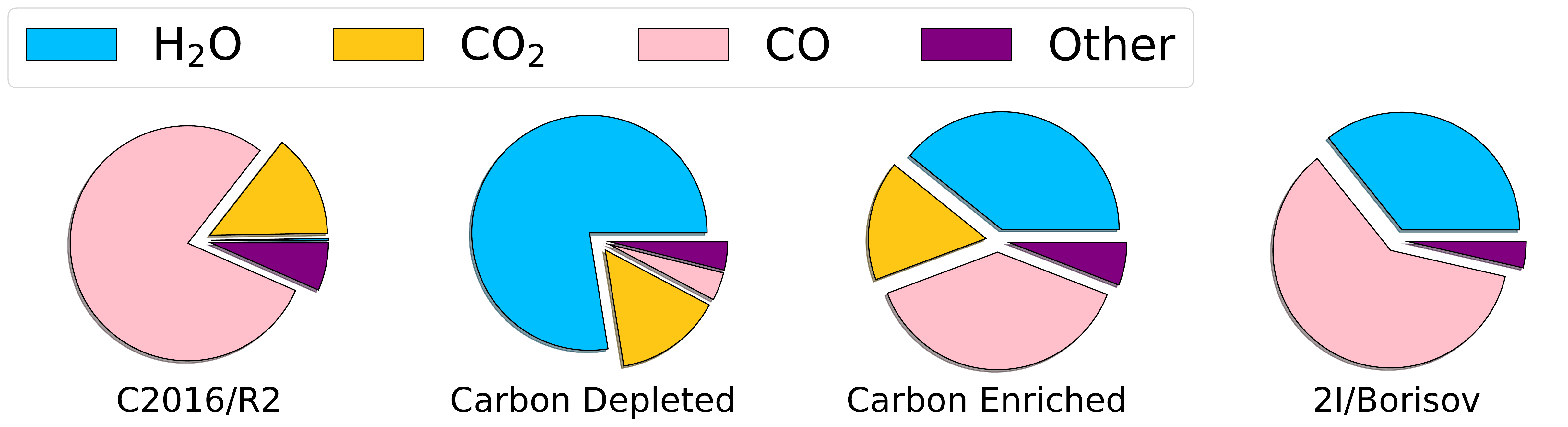}
    \caption{The composition of the LPC C/2016 R2, 2I/Borisov, and  typical carbon enriched and  depleted solar system comets. This is a generalized version of an analogous figure in \citet{McKay2019}, and adapted from \citet{Seligman2022PSJ}. 
   The carbon depleted comet is representative of many of the solar system comets for which production rate measurements of CO$_2$, CO and H$_2$O exist (see Table 1 in \citet{Seligman2022PSJ}).  The carbon enriched comet is W3 Christensen \citep{Ootsubo2012}. The composition for 2I/Borisov is derived from Table \ref{Table:waterco} and the references therein, and  R2 is from \citet{McKay2019}. The lack of CO$_2$ for 2I is only because no measurement of CO$_2$ was reported}\label{fig:piecharts} 
\end{center}
\end{figure}

The $Q({\rm CO}) / Q({\rm H_2O}) \sim$ 1 production rate ratio in 2I/Borisov \citep{Bodewits2020,Cordiner2020} distinguishes this object from most solar system comets, in which  H$_2$O is the dominant molecule (Figure \ref{fig:piecharts}).  The average cometary  ratio is $Q({\rm CO}) / Q({\rm H_2O}) \sim$ 4$\%$, albeit with a wide range from 0.5\% to 20\% \citep{Bockelee22}.  However, a few exceptionally CO-rich comets exist.  For example,  C/1995 O1 (Hale-Bopp) had $Q({\rm CO}) / Q({\rm H_2O}) >$ 12 at $r_H$ = 6 AU \citep{Biver02}, while  short-period comet/Centaur 29P/Schwassmann-Wachmann 1 had $Q({\rm CO}) / Q({\rm H_2O}) =$ 10$\pm$1, also at $r_H \sim$ 6 AU \citep{Biver02,Bockelee22}).  The outstanding example is C/2016 R2, in which $Q({\rm CO}) / Q({\rm H_2O}) = $ 308$\pm$35  at $r_H \sim$ 2.8 AU \citep{McKay2019} (c.f.~$Q({\rm CO}) / Q({\rm H_2O}) > $ 10, \cite{Biver18}).  C/1995 O1 (Hale-Bopp) and C/2016 R2 have barycentric orbital eccentricities $e <$ 1 and are from the Oort cloud, while 29P is a recent arrival from the Kuiper belt.  

How can these large $Q({\rm CO}) / Q({\rm H_2O})$ ratios be understood?  One effect is distance; the $Q({\rm CO}) / Q({\rm H_2O})$  in a given object should naturally grow with $r_H$ because the volatility of water ice falls faster with $r_H$ than that of CO.  This contributes to the high ratios in 29P and C/1995 O1 at 6 AU, where water is largely frozen out.  However, 2I/Borisov and C/2016 R2 were observed at  much more modest heliocentric distances ($\sim$2 to 2.5 AU) and the distance effect can be ignored.  

A second effect is that the outgassed species might not represent the bulk composition of the nucleus.  For example, highly volatile CO molecules can be mobilized at much lower temperatures than H$_2$O molecules, providing a source zone extending more deeply into the nucleus.  This effect must be transient, because CO could be depleted from the thermally heated skin of the nucleus long before the water ice is sublimated away.  But the effect is difficult to model, because cometary volatiles exist within a complex, porous regolith with a large and varying temperature profile and unmeasured permeability to deep gas flow.  

The most interesting interpretation of these observations is that super-volatile enriched objects like 2I/Borisov and C/2016 R2 were accreted at very large stellocentric distances compared to most of the comets remaining in the short- and long-period populations.  Comets formed at large distances beyond the CO snow-line would be the least strongly bound to their parent star and perhaps the most likely to be lost to interstellar space.  If so, we should expect to find higher average $Q({\rm CO}) / Q({\rm H_2O})$ in interstellar interlopers than in the bound comets.

\subsubsection{Evidence from Protoplanetary Disks}

The study of interstellar interlopers is closely related to the field of protostellar disk evolution. Recent observations that resolve protostellar disk substructure are beginning to revolutionize our understanding of the earliest stages of planet formation \citep{Manara2022,Miotello2022}. Observations of face-on disks have revealed the ubiquity of  remarkable structures including gaps, spirals and rings \citep{Long2018,Andrews2018,Andrews2020,Oberg2021,Benisty2022}.  Protostellar gas disks are typically a factor of two larger than disks of millimeter sized dust grains, possibly a result of radial migration of the dust under gas drag \citep{Ansdell2018}. 

Disk observations from near the projected mid-plane allow for the measurement of the vertical scale heights of both gas and dust. Gas is vertically supported by a pressure gradient while, in the absence of substantial turbulence, dust settles towards the mid-plane under the action of stellar gravity.  As a result, the dust mass density in the mid-plane grows towards a critical value. In the modern view, a ``streaming instability'' is triggered if the dust to gas ratio of $\sim1$ is reached \citep{Youdin2005} and the subsequent agglomeration of macroscopic bodies is rapid.  While  volatile molecules can be trapped in water either as clathrates or in the amorphous form, bulk ice can only freeze past the snowline.  Therefore, we are especially interested in measurements of disk structure, including the disk scale-heights in gas and dust, at the largest distances from the parent star.

\begin{figure}
\begin{center}
       \includegraphics[scale=0.3,angle=0]{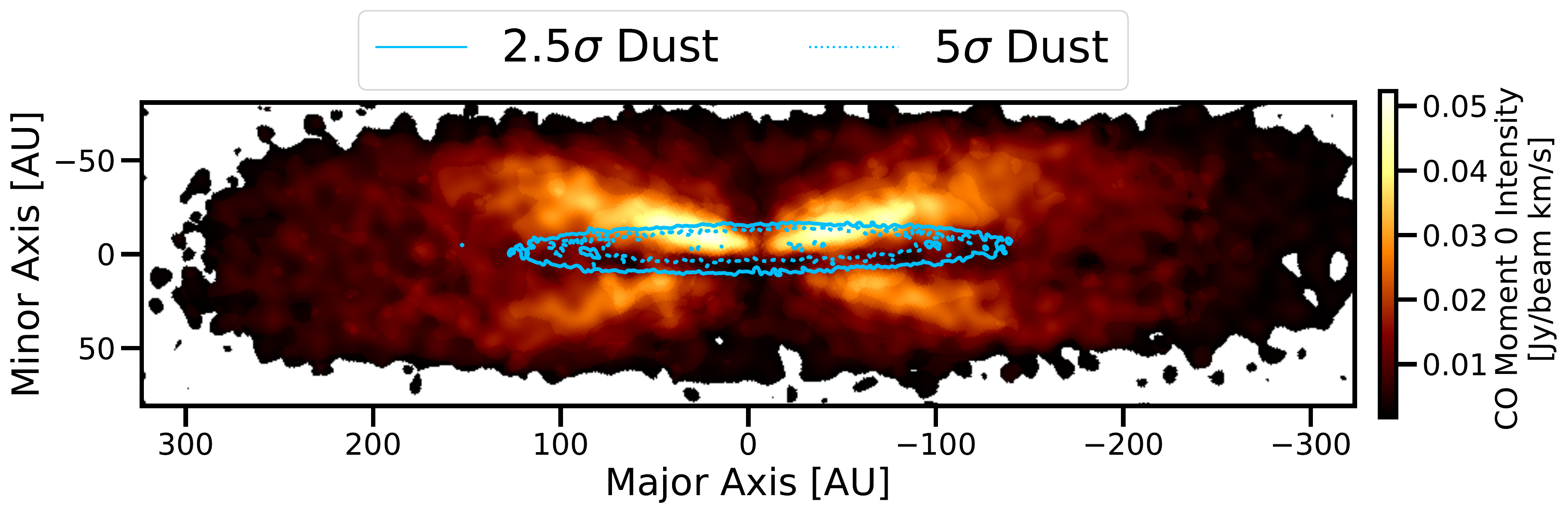}
    \caption{ALMA continuum and CO observations of the edge on protoplanetary disk SSTC2D J163131.2-242627 (Oph163131, $i\sim84^\circ$). The color scale shows the $^{12}$CO  
    intensity map from ALMA observations. The blue solid and dashed lines show $2.5$ and $5\sigma$ contours of the dust continuum from ALMA.  The gas is significantly more extended in the vertical direction than the dust (with scale heights different by more than an order of magnitude), implying conditions amenable for supervolatile enriched planetesimal formation at large stellocentric distances. Adapted from \citet{Villenave2022}. }
 \label{fig:oph} 
\end{center}
\end{figure}

\citet{Villenave2022} presented ALMA images   of the protoplanetary disk SSTC2D J163131.2-242627 (Oph 163131) whose disk is inclined to the line of sight by only $i\sim6^\circ$.  The  scale height in millimeter-sized dust is $\sim$0.5 AU at 100 AU from this Sun-like star (the mass is 1.2$\pm$0.2 M$_{\odot}$), compared with a scale height in $^{12}$CO gas of $\sim$10 AU at the same distance (Figure \ref{fig:oph}). Enhanced mid-plane densities may allow macroscopic bodies to grow rapidly in this disk, even at 50 AU to 100 AU from the star.  SSTC2D thus constitutes a possible analog for the formation site of CO-rich 2I/Borisov \citep{Bodewits2020,Cordiner2020}.  Other works also find a dust scale height of $\sim1$AU at 100AU from the central star \citep{Pinte2016,Villenave2020,Doi2021}.

Given the greater abundance of H$_2$O relative to CO in molecular clouds and protoplanetary disk gas, some mechanism of concentration is needed to account for comets in which  CO/H$_2$O  $\gtrsim$ 1. Radial transport of solids under the action of viscous forces may play a role.  New models including both the migration of solids and the diffusion and freezing of CO gas in a protoplanetary disk show the formation of an extensive region in which grains grow with CO/H$_2$O $>$ 1 \citep{Price21,Estrada22}. These models are highly idealised,  involving many assumptions and poorly characterized physical processes.  However, the growth of a CO-enriched region, spanning from about 10 AU to 100 AU after 10$^6$ years of disk evolution, appears to be a consistent result. Such regions could be the source of CO-enhanced comets and interstellar objects.

Perversely, the above models predict CO enrichment over such a large fraction  of the protoplanetary disk that one wonders why CO-enriched comets are observationally rare.  A possible answer is given by \citet{Lisse2022}, who argued that the final super-volatile abundances of comets should also depend  on their disk residence time prior to ejection from the parent star. For example, \citet{Steckloff2021} argued that the solar radiation received by Arrokoth and other Cold Classical KBOs, when integrated over $>10-100$ million year timescales, would be sufficient to deplete all subsurface super-volatiles. Comets ejected to interstellar space soon after their growth can preserve a high CO fraction, while those lingering longer would lose the bulk of their super-volatiles by sublimation.   In this view, the low $Q({\rm CO}) / Q({\rm H_2 O})$ in most measured comets reflects a long interval between the growth of the nucleus and the formation of the Oort cloud.  CO-enriched objects like C/2016 R2 and 2I/Borisov would then have preserved their CO by virtue of unusually early ejection from the parent disks. The apparent rarity of CO-enriched comets may simply reflect the time profile of cometary ejection.

 Unfortunately, even in the solar system, the details of cometary accretion are subjects of uncertainty and contention \citep{Davidsson21,Davidsson2022} and the timing of Oort cloud formation is essentially unknown, making these ideas  difficult to test.  Still, it is evident that future measurement of the distribution of compositions of interstellar interlopers will provide insights into both the timing and the structure and evolution of the disks from which they are likely ejected.

\section{Statistics and Origins}

\subsection{Galactic population} 
\label{sec:limits}

Prior to the discovery of 1I and 2I, only upper limits to the galactic number density of interstellar bodies could be estimated.  These estimates are scattered over a wide range, depending on assumptions made about the observational parameters of different surveys, and about the nature and distribution of the sources of interlopers. The most ``sophisticated'' estimates were not necessarily the most accurate.  
   
The detection of `Oumuamua  in the relatively well-characterized Pan STARRS sky survey allows for more confident estimates of the galactic number density of similar bodies.  The value is now estimated to be between $n_{o} \sim$ 0.1 AU$^{-3}$ \citep{Jewitt2017, Trilling2017} to 0.2 AU$^{-3}$  \citep{Do2018}.  Since 5 years have elapsed since the discovery of `Oumuamua without another detection of a similar object, we adopt the lower estimate, $n_{o}\sim 0.1$ AU$^{-3}$.  This corresponds to $\sim$ 10$^4$ similar objects closer to the Sun than Neptune (i.e.~distance $\le$ 30 AU) at any instant.  With a solar system crossing time $\sim$ 10 years, the flux of interlopers into the planetary region is an incredible $\sim$ 10$^3$ year$^{-1}$ (3 day$^{-1}$).

Considering the galaxy as a disk of radius 10 kpc and thickness 1 kpc, this density implies a population $\sim$ 10$^{26}$ objects of 100 m scale, with a combined mass $\sim$6$\times10^{35}$ kg ($10^{11}$ M$_{\oplus}$, or about 1 M$_{\oplus}$ per star). This is comparable to the canonical $\sim$1 M$_{\oplus}$ estimated mass of the Oort cloud, but refers to objects 10 times smaller.    However, the  interloper mass is very uncertain because it is based on a single object and because interstellar objects presumably occupy a size distribution in which the mass is dominated by objects larger than `Oumuamua.  Given these uncertainties, it is not yet clear if the inferred interloper population contradicts the hypothesis that these objects are ejected comets.

Conversely, 2I/Borisov was discovered as part of a near-Sun survey whose depth and areal coverage have not been published, making it impossible to estimate a useful number density.  Furthermore, 2I  was discovered because of its bright coma, the uncertain optical properties of which would undercut any attempt to derive a meaningful object number density. For these reasons, it is not possible to use the detection of the second interloper to strengthen the density estimate obtained from `Oumuamua.

\subsection{Dynamics}

\begin{figure}
\begin{center}
       \includegraphics[scale=0.32,angle=0]{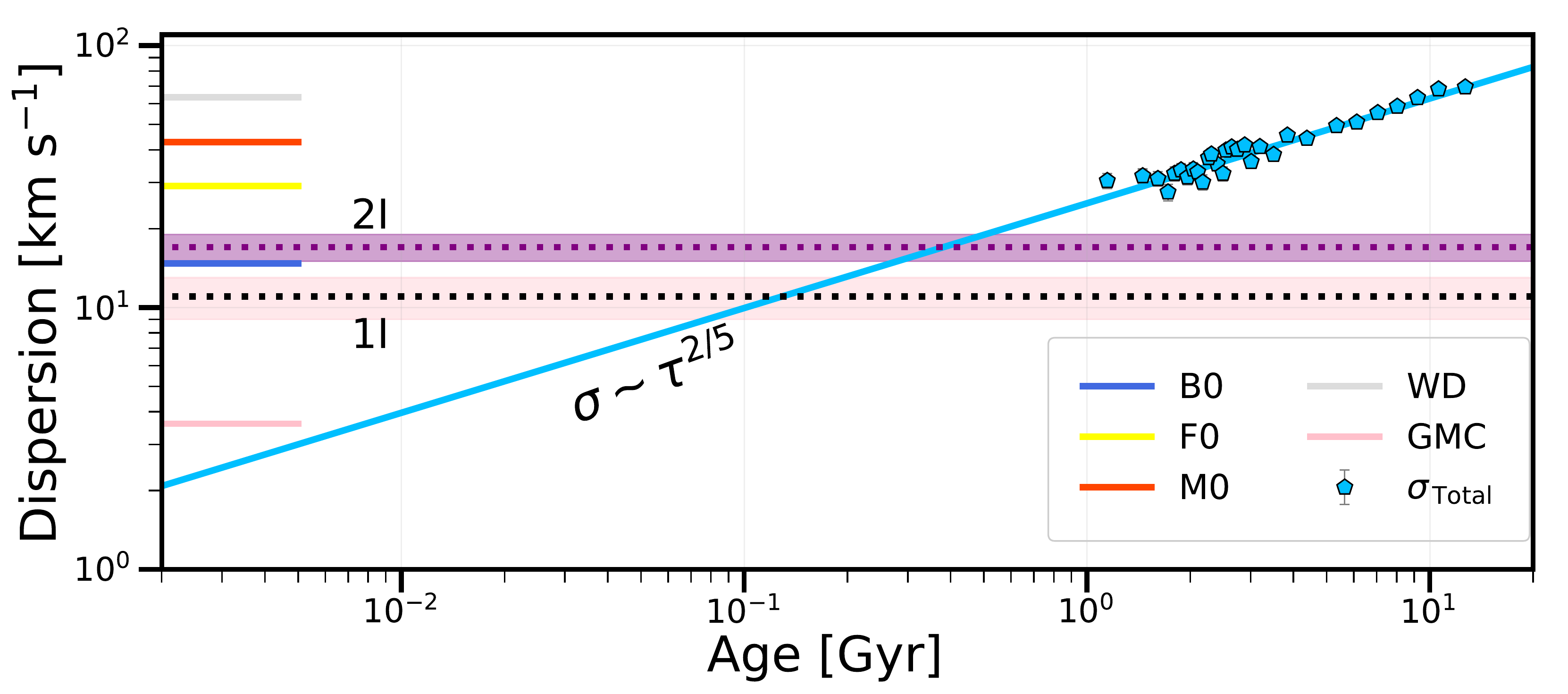}
    \caption{Age kinematics of stars and interstellar interlopers. Blue pentagonal points show average measured stellar velocity dispersions (Figure 7 in \citet{Holmberg09}), with a best fit $\sigma\sim \tau^{2/5}$ as a blue solid line. The  black and purple dotted lines indicate the  velocity for 1I/`Oumuamua and 2I/Borisov (with only statistical uncertainties shaded). Median measured  dispersions of stars, White Dwarfs \citep{BinneyMerrifield1998}  and GMCs \citep{MivilleDeschenes2017} are indicated with solid lines.
    }\label{fig:age} 
\end{center}
\end{figure}

Gravitational interactions with giant molecular clouds (GMCs) and other sub-structures in the disk of the galaxy cause a progressive excitation of the velocities  of passing stars and tracer particles.  This process of ``disk heating'' should similarly excite the motions of interstellar bodies, providing a method to estimate the length of time they have spent in the interstellar environment. In fact, since  interstellar objects outnumber  stars by many orders of magnitude, the population is in principle a much \textit{better} realization of the fine-grain assumption in the collisionless Boltzmann equation. Figure \ref{fig:age} shows an empirical stellar age vs.~velocity dispersion relation  together with the excess velocities relative to the local standard of rest of both interlopers \citep{Holmberg09}.     `Oumuamua's low  velocity (26 km s$^{-1}$, compared to 15$\pm$2 km s$^{-1}$ for the velocity of the Sun relative to LSR \citep{Robin17}) implies  an age $\tau_s \sim$ 100 Myr, originally noted by \cite{Mamajek2017, Gaidos2017} and \citet{Hallatt2020}.  The larger velocity of 2I/Borisov (32 km s$^{-1}$),  indicates greater disk heating and therefore a greater age since ejection, probably $\tau \sim10^9$ yr \citep{Hallatt2020}. These estimates are statistical in nature and also subject to surprisingly large systematic uncertainties in the velocity of the Sun (\cite{Schonrich12} vs.~\cite{Robin17}); they cannot be used to infer highly accurate ages. Still, the strong likelihood is that 1I/`Oumuamua has spent less time in interstellar space than 2I/Borisov.     

 With an age $\sim$100 Myr, `Oumuamua has probably not travelled far from its origin.
 \citet{Gaidos2017}  identified a  match with the $45\pm10$ Myr old Carina and Columba moving groups (co-moving but unbound aggregates of recently formed stars) \citep{Bell15}. He suggested that `Oumuamua formed in a protostellar disk around a star there and was  ejected  with low peculiar velocity to explain the kinematics.  Independent dynamical integrations of the galactic trajectory confirm that `Oumuamua was very likely associated with the local Orion Arm, consistent with the Carina or Columbia stellar associations \citep{Hallatt2020}. \citet{Hsieh2021} argued that this was evidence that the object was  produced in a giant molecular cloud core, instead of a protostellar disk - because these cloud products would have significantly lower velocity dispersions due to the transient nature of star forming regions in the galaxy. 
While linkage to a general region of the galaxy is possible, attempts to identify the particular star from which `Oumuamua was ejected are futile, given the many observational and dynamical uncertainties.  There is even less hope of identifying a home system for the much older 2I/Borisov \citep{Hallatt2020}. 

One consequence of a local origin for `Oumuamua is that the inferred density of similar objects, $n_0 \sim$ 0.1 AU$^{-3}$, may not apply uniformly to the whole galaxy. \citet{MoroMartin2018i} argued that `Oumuamua was ejected from the planetesimal disk of a young nearby star, and that this ejection was highly anisotropic.  \citet{Hsieh2021}  traced the motions of test particles ejected from stars in the Carina and Columba stellar associations and found them to be statistically consistent with the orbit of 'Oumuamua.  They inferred ejection speeds $\sim$1 km s$^{-1}$.

\subsection{Effects of the interstellar environment on the interlopers}
Direct interactions between stars and interlopers are extraordinarily rare. The timescale for a single object to pass within a distance, $d$, of a star is just $t \sim (\pi N_{\star} \Delta V d^2)^{-1}$, where $N_{\star} \sim$ 0.1 pc$^{-3}$ is the number density of stars, and $\Delta V$ = 50 km s$^{-1}$ is the nominal velocity dispersion.  
The paths of interlopers will be affected by gravitational focusing, which increases the effective cross-section by a factor  $ \sim  (V/\Delta V)^2$, where  $V$ is the local escape speed at the minimum distance from the star. For example, a Sun impact would have $d \sim 10^9$ m and $V/\Delta V \sim (600/50)^2 \sim$ 140. Substituting, we find $t \sim 10^{17}$ years, showing that star-interloper collisions  can be ignored. Even interactions as close as those of `Oumuamua (minimum distance $d$ = 0.25 AU) and 2I/Borisov ($d$ = 2.0 AU) are incredibly unlikely ($t \sim 10^7$ Gyr), for a given object. As a result, the pre-entry thermal evolution of interstellar interlopers should be minimal. Their surface temperatures, set by equilibrium with the interstellar UV flux, will be just a few degrees above the microwave background temperature.

On the other hand, the interlopers travel through and interact with the gas and dust of the interstellar medium.  \cite{Stern1990} showed that while some mass is added by the implantation of interstellar gas  into the surfaces of Oort cloud comets (which are as fully exposed to the ISM as are the interlopers), much more material is eroded by impact with interstellar dust grains.  The net effect is the loss of the upper $\Delta L \sim 0.1$ m of surface for every billion years of exposure.  Compared to the size scales of `Oumuamua and 2I/Borisov, the lost material is mass-wise unimportant.  Only decimeter scale and smaller interstellar debris should be substantially depleted by impact erosion on billion year timescales.  

Interstellar space is also pervaded by cosmic rays,  with energies that are orders of magnitude larger than the few eV binding energies of common molecular bonds.  Cosmic rays severely damage the molecular structure of materials with which they interact.  In the open interstellar medium, the cosmic ray energy spectrum  resembles a broken power law, having the largest fluences at the smallest energies. (The energy spectrum in the Kuiper belt is different, owing to shielding of low energy particles by the Sun's magnetic field). The cosmic ray penetration depths into solid matter, $d_{CR}$ depend on the particle energies, $E$. For  $E \lesssim$ 0.1 GeV the penetration depths, $d_{CR} <$ 0.1 m \citep{Cooper03,Gronoff20}, are smaller than the impact-eroded layer thickness.  With $d_{CR} < \Delta L$,   the degree of surface processing by low energy particles is limited by the steady loss of surface material to impact and the continual exposure of fresh material from beneath.  However,  the energetic $E \gtrsim$ 1 GeV protons and alpha particles have  much larger penetration depths in ice,  $d_{CR} \sim$1 m to 10 m.  Since  $d_{CR} > \Delta L$,  comets and interstellar interlopers should develop a processed surface layer or ``crust'' that is too thick to be eroded away by impact with interstellar dust.  

Laboratory experiments show that the principal effect of energetic particle bombardment is to sever molecular bonds, resulting in the formation of new bonds (and radicals). Being light, hydrogen atoms easily escape, leading to a progressive build-up of macro-molecular, carbon-rich solids. This irradiated material can have low volatility and very low albedo, quite different from the initial material.  An unfortunate consequence of the destruction of bonds is that irradiated materials lack the characteristic vibrational spectral features on which spectroscopic chemical identifications are based (e.g.~\cite{Faure21}).  Probably for this reason, the reflection spectra of most comets, like those of `Oumuamua and 2I/Borisov, tend to be linear, featureless and difficult or impossible to compositionally diagnose.  The similarity between the redder-than-sunlight colors of the interlopers and those of solar system comets (Section \ref{colors})  is broadly consistent with  cosmic ray processing of both.

\subsection{Capture of Interstellar Objects }

Galactic tides, although very weak, can temporarily trap slowly passing interstellar objects on loosely bound orbits, building a swarm of such bodies estimated to number $\sim10^7$ \citep{Penarrubia22}.  Likewise, a small fraction of the interstellar objects passing through the planetary region of the solar system can  be trapped by gravitational interactions with the planets, albeit with very low and strongly encounter-velocity dependent efficiency
 \citep{Napier2021,Dehnen2022I}.  The action of non-gravitational acceleration could also trap smaller objects.
 
 \citet{Napier2021b} calculated that the total mass of interstellar objects trapped in the Solar System was $\simeq10^{-9}M_\oplus$, by estimating their typical dynamical lifetimes. Most of this interstellar material in the present day Solar System was captured during the Sun's cluster phase. Under a different set of assumptions, \citet{Dehnen2022II} argued that there were only $\sim 8$ interstellar objects captured within 5AU at any given time.

While the statistics of capture are forbidding, it has nevertheless been suggested that some Centaurs and Trojans with extreme orbits could be captured interstellar bodies \citep{Namouni20}.  However,  it is more likely that these objects are  transient captures from distant solar system reservoirs, such as Halley type comets or the Oort cloud \citep{Morbidelli2020}.

In this regard, just as there is a small probability for unbound objects to become trapped into bound orbits, it is possible for comets from our own Oort cloud to be scattered onto  hyperbolic orbits through interactions with passing stars and sub-stellar objects.  On rare occasions,  ejected Oort cloud comets passing through the planetary region might be mistaken for interstellar interlopers arriving from afar. \cite{Higuchi20} estimated that about 0.1\% of objects with orbits similar to `Oumuamua, and 0.01\% of those similar to the more eccentric orbit of 2I/Borisov, could originate as comets deflected from our own Oort cloud.  While these probabilities depend on poorly constrained estimates of the number density of nearby sub-stellar and even sub-Jovian perturbers, it is clear that almost all unbound objects entering the planetary region originate from elsewhere in the galaxy.

\subsection{Interstellar Meteors and Impactors}
\label{meteors}

\begin{figure}
\begin{center}
       \includegraphics[scale=0.28,angle=0]{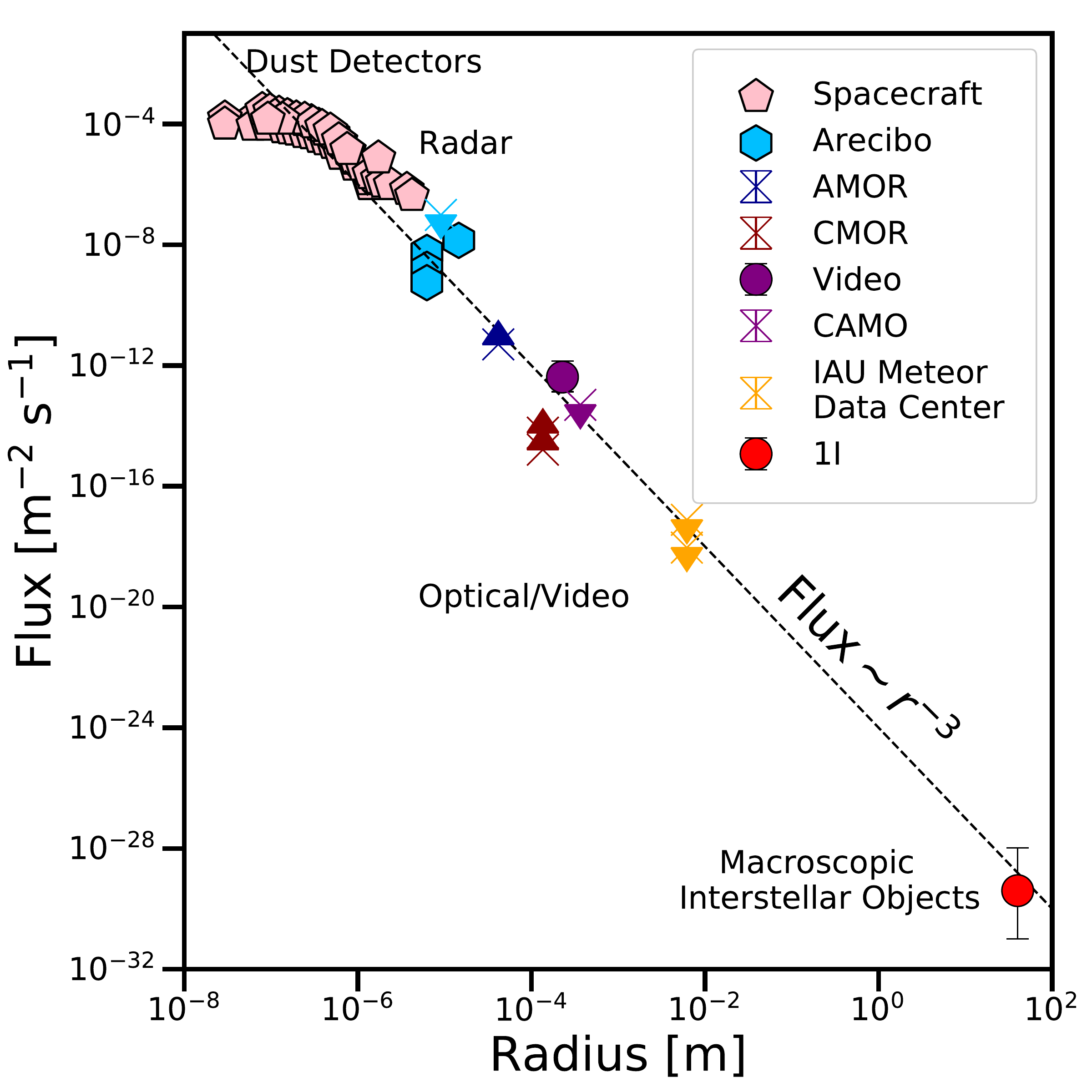}
    \caption{Observational constraints on the fluxes of interstellar bodies. The line is added to guide the eye and is not a fit to the data.   Figure modified from \citet{Musci12}. 
      } 
 \label{fig:musci} 
\end{center}
\end{figure}

Existing constraints on the flux of interstellar particles across a wide range of sizes are shown in Figure \ref{fig:musci}.  The observations consist of (pink pentagons) in-situ dust detections from the Ulyssess and Galileo spacecraft \citep{Grun1993,Grun1997,Landgraf1998,Grun2000,Landgraf2000}, (blue hexagons) radar measurements from Arecibo \citep{Mathews1999,Meisel2002a,Meisel2002b}, radar lower limits with the  (dark blue X) Advanced Meteor Orbit Radar (AMOR) \citep{Baggaley1993,Baggaley2000,Taylor1996} and (red X) Canadian Meteor Orbit Radar (CMOR)\citep{Weryk2004}, (purple triangle) optical data from  \citet{Hawkes1999} and  from  the Canadian Automated Meteor Observatory (CAMO) \citep{Musci12}, and (orange X) upper limits from optical images of meteoroids from the photographic database of the IAU Meteor Data Center \citep{Hajdukova1994,Hajdukova2002}.  The flux inferred from the 0.1 AU$^{-3}$ density of 1I/`Oumuamua-like objects is shown as a red filled circle.

The data in Figure \ref{fig:musci} are broadly compatible with a $r_n^{-3}$ radius distribution, subject to a roll-over at sizes $r_n \lesssim$0.1 $\mu$m. The latter occurs because dust particles smaller than $\sim$0.1 $\mu$m are largely deflected from the inner solar system by a combination of radiation pressure and Lorenz forces. Slightly larger  particles (radius $\gtrsim$0.4 $\mu$m) can penetrate and have been recorded from impact detectors on spacecraft, with a flux $\sim10^{-4}$ m$^{-2}$ s$^{-1}$ \citep{Grun94} (see Figure \ref{fig:musci}).  These interstellar dust particles can be reliably identified because their velocities are  accurately measured using time-of-flight detectors and found to be higher than the local solar system escape speed.

Millimeter-sized interstellar particles should, in principle, give rise to meteors, which should also be readily identifiable by their velocities\footnote{While $e >$ 1 is a necessary condition for the identification of interstellar material, it is not definitive, because planetary perturbations can generate hyperbolic trajectories, albeit with excess velocities of only $\sim$0.1 km s$^{-1}$ \citep{Wiegert2014,Higuchi20}}. However, rather stringent measurements of speed (accurate to $\pm$1 km s$^{-1}$) and of direction (to $\pm$1$^o$ or 2$^o$) are needed to distinguish some hyperbolic from bound orbits  (c.f.~\cite{Hajdukova2020}.  Unfortunately, measurements of sufficient accuracy are difficult given the short flight lengths (a few $\times$10 km) and times ($\sim$1 s) of meteors in the atmosphere.   As a result, there is a long history of false detections of interstellar meteors.  Famously, \cite{Opik34} used optical data to infer high speeds requiring a substantial fraction of terrestrial meteors to be of interstellar origin, a result that we now know to be wholly incorrect.  Similarly, \cite{Siraj19}  claimed that a previously reported  meteor was interstellar but relied, in part, on secret (US Dept. Defense) data concerning the accuracy of the trajectory.  Interferometric radar determinations of meteor velocities \citep{Baggaley07} show that most meteors are unambiguously bound and thus have a solar system origin, with only a small high velocity tail, again likely due to measurement errors \citep{Musci12}.

Large interstellar objects must occasionally strike the Earth but, because of their high impact velocities, they are less likely than asteroids to deposit meteorites onto the surface.   The rate of  impact of 100 m scale interstellar bodies into Earth is $\sim$(5 to 10)$\times10^{-9}$ year$^{-1}$ \citep{Jewitt20}, giving only 25 to 50 such events over the age of the Earth. It is possible  that a majority of these would have exploded above the ground as airbursts.  This estimated rate is $\sim10^{4}$ times less than the rate of impact by solar system material (mostly asteroids) of comparable size. Therefore, we are unlikely to find evidence  for craters formed by interstellar projectiles or for interstellar meteorites in our collections.  Even if the flux were higher, it is unclear how to distinguish a crater formed by an interstellar impact from one formed by a solar system projectile  \citep{Cabot2022}.

The mass distribution of free-floating planets resembles a differential power law
having an index $p$ = 0.92$\pm$0.06 \citep{Gould22}.  This corresponds to a size distribution index 3.8$\pm$0.2, which is comparable to but slightly steeper than suggested by Figure \ref{fig:musci}.  While this similarity is likely a coincidence, as noted by \cite{Gould22},  it is also an interesting reminder that young planetary systems can eject objects with size scales ranging from dust to planets.

\section{Future Prospects}

\subsection{Ground-Based}

It is  possible that  interstellar objects in the solar system have been recorded but went unnoticed in existing survey data. A systematic search for such objects in archival data would be a valuable first step towards improving our population estimates, even before the advent of powerful, new sky survey telescopes. Many new interstellar interlopers are expected to be found as products of all-sky surveys both planned and already  operational. In particular, the Rubin Observatory Legacy Survey of Space and Time (LSST) should offer a substantial increase in sensitivity to transient objects, while surveying the entire night sky in the southern hemisphere with a close to nightly cadence  \citep{Moro2009,Engelhardt2014,Cook2016,Seligman2018}. \citet{Hoover2022} used an elaborate simulation to estimate not only the LSST detection rates as a function of size and absolute magnitude, but also the distribution of orbital elements and trajectories of detectable interstellar objects. They predicted that the LSST will detect between 1-2 `Oumuamua-like interstellar objects every year (for details see Table \ref{tab:percent_table}).

\begin{table}
\tabcolsep1.5pt
\caption{Predicted detection rates of interstellar objects$^a$ adapted from \citet{Hoover2022}. }
\label{tab:percent_table}
\begin{center}
\begin{tabular}{|@{}c|c|c|c|@{}}
\hline
Criterion & Percent& Conservative Rate per Year&Optimistic Rate per Year\\\hline 
    Detectable by the LSST & 2.828\% & $\sim 0.9$ & $\sim 1.9$\\
    \hline
    Detectable, $\Delta V < 15$ km/s & 0.424\% & $\sim 0.1$ & $\sim 0.3$\\
    \hline
    Detectable, $\Delta V < 2$ km/s & 0.002\% & $\sim 0.0007$ & $\sim 0.001$\\
    \hline
\end{tabular}
\end{center}
\begin{tabnote}
$^{\rm a}$ Assuming `Oumuamua-like absolute magnitude, H = 22.4. From \citep{Hoover2022} \\
\end{tabnote}
\end{table}

The flux curve in Figure \ref{fig:musci} indicates that meter-sized interstellar objects, with a flux $f \sim 10^{-24}$ m$^{-2}$ s$^{-1}$, should strike the Earth's atmosphere on a $t \sim (4\pi R_{\oplus}^2 f)^{-1} \sim10^2$ year timescale.  It is therefore not surprising that convincing examples of hyperbolic bolides have yet to be reported. However, the cumulative fluxes of 1 cm and 10 cm scale interstellar meteors should be higher by 4 and 2 orders of magnitude, respectively, corresponding to timescales from a few days to a year.  These timescales approach the duration of long-term meteor surveys, such that we can reasonably hope for convincing detections of small scale interstellar meteors in the next few years, provided that adequate velocity resolution can be attained.   Firm detections of material in the centimeter to decimeter size range would help bridge the gap between the spacecraft detected micron-sized interstellar dust particles and the macroscopic objects discussed here.

\subsection{Space Based}
In space, the forthcoming NEO Surveyor (consisting of a 50 cm diameter telescope located interior to Earth's orbit at  L1 (\cite{Mainzer2015}) is expected to  provide thermal infrared (10 $\mu$m) detections and orbits of small bodies out to the orbit of Jupiter, with a sensitivity to near-Earth objects rivalling that of the LSST.  Located interior to Earth's orbit, it will also provide greater coverage of the sky at small elongation (Sun-telescope-object) angles than is possible with a large telescope from the ground. This allows for the detection of objects at heliocentric distances $r_H <$ 1 AU.  Combined with optical data, NEO Surveyor thermal flux density measurements  will break the degeneracy between the sizes and albedos that afflicts optical data alone.  Thermal data from JWST (\textit{James Webb Space Telescope}) can also fulfill this role, albeit with much more restricted telescope pointing constraints.

Interstellar comet analogues might also provide a source of volatile enrichment to short period planets. The discovery of the first  interstellar object implies that, on average, every star contributes $\sim 1 M_\oplus$ worth of cometary material to the galactic population.  If every star ejects $\sim 1 M_\oplus$  of  material and because  ejection is a chaotic process, a comparable amount of material should be injected into the interior of every contributing system and potentially accreted by exoplanets \citep{Seligman2022ApJL}. This enrichment could constitute a non-negligible fraction of the atmospheric metal content of  exoplanetary atmospheres.

Recently, the European Space Agency selected the \textit{Comet Interceptor} \citep{jones2019} to launch in 2029. The Interceptor has a low $\Delta V \sim$ 1 km s$^{-1}$ 
 budget and can only reach objects whose orbits bring them fortuitously close to its loitering location at L2 within the 3 year duration of its mission \citep{Sanchez21}.  With these mission parameters, the likelihood that Interceptor will find an accessible interstellar target is negligible (c.f.~Table \ref{tab:percent_table}, \cite{Hoover2022}). More optimistically, an impactor mission to `Oumuamua, sent from the Earth, would have been achievable with a modest impulse ($\Delta V \sim$ 4 km s$^{-1}$) given sufficient forewarning of the approach \citep{Seligman2018}.  \citet{Hoover2022} estimated that 10\% to 30\% of  the interstellar interlopers to be detected by the LSST  will be reachable by a mission with $\Delta$V $<$ 15 km s$^{-1}$ \citep{Hoover2022}. Optimistically,  a handful of rendezvous-suitable targets will be detected each decade.

\begin{summary}[SUMMARY POINTS]
\begin{enumerate}
\item The  two known interstellar interlopers are both sub-kilometer bodies exhibiting non-gravitational acceleration but otherwise are surprisingly physically different. 1I/`Oumuamua appears asteroidal while 2I/Borisov outgasses strongly. It is not clear whether these differences reflect two different populations of interstellar body or different evolutionary stages of the same  type of object.

\item Dynamical considerations suggest that 1I/`Oumuamua and 2I/Borisov have different ages, likely $\sim10^8$ years and $\sim10^9$ years, respectively.  

\item Interlopers formed by accretion in protoplanetary disks are ejected to  interstellar space by strong gravitational scattering from planets.  `Oumuamua is tentatively associated with the Carina or Columbia stellar associations.  The origin of 2I/Borisov is unknown. 

\item The number density of `Oumuamua-like (100 m scale) bodies is $\sim$0.1 AU$^{-3}$, and the implied galactic population is $\sim$10$^{14}$ to 10$^{15}$ per star.  
\end{enumerate}
\end{summary}

\begin{issues}[FUTURE ISSUES]
\begin{enumerate}
\item Current interloper population estimates are extremely uncertain but of great importance, in relation to the likely protoplanetary disk origins of these bodies and their galactic total mass.
\item Physical measurements of interstellar interlopers are needed to understand the reason for the divergent appearances of the first two examples, and to better relate these bodies to solar system comets.
\item Planned deep, all-sky surveys (in particular, by NASA's NEO Surveyor and by the LSST) are expected to reveal $\sim$1 new interstellar interloper per year.
\item Spacecraft intercepts of interlopers will be possible but difficult, given the high average encounter velocities and limited forewarning of arrival.
\end{enumerate}
\end{issues}

\section*{DISCLOSURE STATEMENT}
The authors are not aware of any affiliations, memberships, funding, or financial holdings that
might be perceived as affecting the objectivity of this review. 

\section*{ACKNOWLEDGMENTS}

We thank Marion Villenave, Karen Meech and Olivier Hainaut for providing us with data and Robert Jedicke, Davide Farnocchia, Yoonyoung Kim, Marco Micheli, Greg Laughlin, Aster Taylor, Jane Luu, Pedro Lacerda, Amy Mainzer, Adina Feinstein, Andrew Youdin, Jing Li,  Benjamin Donitz and Alan Stern for  useful conversations and suggestions. 

\bibliographystyle{ar-style2}
    \bibliography{bibfile2}

\end{document}